\documentclass{article}

\usepackage{microtype}
\usepackage{graphicx}
\usepackage{subfigure}
\usepackage{booktabs} 

\usepackage{amsmath}

\usepackage{amssymb}
\usepackage{enumitem}
\usepackage{seqsplit}
\usepackage{xcolor}
\usepackage{xspace}
\usepackage{multirow}
\definecolor{lightgreen}{RGB}{144,238,144}
\definecolor{darkcyan}{HTML}{008B8B}

\newcommand{\sys}{\textsc{HexiScale}\xspace}
\newcommand{\ryan}[1]{{\color{black}{#1}}}
\PassOptionsToPackage{hyphens}{url}\usepackage{hyperref}

\usepackage[accepted]{mlsys2025}

\mlsystitlerunning{\sys: Facilitating Large Language Model Training over Heterogeneous Hardware}

\begin{document}

\twocolumn[
\mlsystitle{\sys: Facilitating Large Language Model \\ Training over Heterogeneous Hardware}



\mlsyssetsymbol{equal}{*}

\begin{mlsysauthorlist}
\mlsysauthor{Ran Yan}{equal,to}
\mlsysauthor{Youhe Jiang}{equal,to}
\mlsysauthor{Xiaonan Nie}{pku}
\mlsysauthor{Fangcheng Fu}{stju}
\mlsysauthor{Bin Cui}{pku}
\mlsysauthor{Binhang Yuan}{to}
\end{mlsysauthorlist}

\mlsysaffiliation{to}{Department of Computer Science and Engineering, The Hong Kong University of Science and Technology, Hong Kong, China}
\mlsysaffiliation{pku}{School of Computer Science, Peking University, Beijing, China}
\mlsysaffiliation{stju}{School of Artificial Intelligence, Shanghai Jiao Tong University, Shanghai, China}


\mlsyscorrespondingauthor{Binhang Yuan}{biyuan@ust.hk}

\mlsyskeywords{Machine Learning, MLSys}

\vskip 0.3in

\begin{abstract}
Training large language models (LLMs) is a computationally intensive task, which is typically conducted in data centers with homogeneous high-performance GPUs. In this paper, we explore an alternative approach by deploying training computations across heterogeneous GPUs to enable better flexibility and efficiency for heterogeneous resource utilization. Toward this end, we propose a novel system, \sys, that can flexibly support asymmetric partition of training computations in the scope of data-, pipeline-, and tensor model parallelism. We further formalize the allocation of asymmetric partitioned training computations over a set of heterogeneous GPUs as a constrained optimization problem and propose an efficient hierarchical graph partitioning algorithm. Our approach effectively allocates training computations across heterogeneous GPUs, fully leveraging the available computational power. We compare the performance of \sys with state-of-the-art homogeneous and heterogeneous training systems. When training LLMs at different scales (from 7B to 30B), empirical results demonstrate that: (\underline{i}) compared to state-of-the-art homogeneous baselines running over homogeneous GPUs, \sys achieves \textit{similar} performance when running over heterogeneous GPUs with the \textit{same} theoretical FLOPS; (\underline{ii}) compared to state-of-the-art heterogeneous baselines running on the same heterogeneous clusters, \sys delivers $1.5\times$ to $2.4\times$ higher throughput.
\end{abstract}]



\printAffiliationsAndNotice{\mlsysEqualContribution} 

\section{Introduction}
Over the past few years, large language models (LLMs) have demonstrated impressive performance and sparked a new wave of exciting AI applications~\cite{bommasani2021opportunities}. However, training these LLMs, such as GPT~\cite{gpt4o}, Claude~\cite{claude3}, Gemini~\cite{reid2024gemini}, Llama~\cite{touvron2023llama,dubey2024llama}, Mixtral~\cite{jiang2024mixtral}, Yi~\cite{young2024yi}, Falcon~\cite{falcon180b}, etc., can be extremely computation-intensive, often involving thousands of GPUs running for months. The high cost of deploying such training tasks in a cluster with homogeneous GPUs has become an obvious obstacle limiting the evolution of LLMs. We explore an alternative approach by \textit{distributing training computations across heterogeneous GPUs}, enabling greater flexibility in resource utilization and democratizing LLM training.


Distributing parallel training computations across heterogeneous GPUs is a natural option to democratize LLM training. In the current exciting era of generative AI, chip vendors typically release new generations of AI chips every 24 months. For instance, Nvidia introduced the Turing architecture in 2018~\cite{Nvida_turing}, Ampere in 2020~\cite{Nvida_ampere}, Hopper in 2022~\cite{Nvida_hopper}, and Blackwell is scheduled for Q4, 2024~\cite{Nvida_blackwell}. On the other hand, one particular version of an AI chip often remains in use by cloud service platforms, technology companies, or research institutions for a much longer period. For example, K80 GPUs with Tesla architecture, released in 2006~\cite{Nvidia_tesla}, are still available on AWS as \texttt{p2} instances~\cite{Amazon}. This observation highlights the important opportunity to explore effective ways to maximize the efficiency of such widely available yet heterogeneous hardware to facilitate more cost-effective and accessible LLM training.

On the other hand, deploying large-scale training computation for LLM over a set of heterogeneous GPUs with different technique specs would be a challenging task regarding training system design and implementation. To effectively distribute the training computation over thousands of GPUs, the state-of-the-art training systems, like Megatron~\cite{narayanan2021efficient} and DeepSpeed~\cite{rajbhandari2020zero}
usually supports: (\underline{i}) tensor model parallelism~\cite{narayanan2021efficient,nagrecha2021model}; (\underline{ii}) pipeline parallelism~\cite{huang2019gpipe,narayanan2019pipedream,yang2021pipemare,narayanan2021memory}; and (\underline{iii}) data parallelism (with potentially sharded implementations of parameters, gradients, and optimizer states across multiple devices, also known as fully sharded data parallelism) ~\cite{rajbhandari2020zero,ren2021zero,rajbhandari2021zero,jiang2023osdp}. However, these systems typically only support homogeneous configurations, which require the entire training cluster to operate under a fully symmetric setup. This means all tensor model parallel groups must have the same degree of parallelism, and the same applies to pipeline parallel groups, as well as data or optimizer parallel groups. Such implementation assumes all the GPUs take the \textit{same} amount of computation load, which significantly limits the system efficiency when deploying the training computation over GPUs with \textit{different} computation capability (measured by the peak FLOPS), \textit{different} device memory (i.e., HBM) capacity, and \textit{different} network bandwidth for each pair of GPUs (inter- and intra-node).

Concretely, there are two fundamental challenges stemming from the \textit{heterogeneity}:

\begin{itemize}[leftmargin=*]
\vspace{-1em}
    \item \textbf{Different GPU computation capability and memory capacity.} Heterogeneous GPUs can vary significantly in terms of computation capability (i.e., FLOPS) and memory capacity. This disparity poses a challenge in distributing computations across all available resources. Without proper resource management, the most capable GPUs may remain underutilized while less powerful GPUs become performance bottlenecks, resulting in system inefficiencies and prolonged training duration. Therefore, partitioning computations to match the capabilities of each GPU is essential to fully utilize the available hardware.
    \vspace{-0.25em}
    \item \textbf{Different GPU-GPU network bandwidth.} The heterogeneous connections between GPUs, ranging from high-speed NVLink and PCIe to standard Ethernet, add another layer of complexity. These varying connection speeds result in uneven communication times. Effective management of communication overhead is essential to prevent faster connections from stalling, waiting for slower ones. This requires sophisticated optimization algorithms to align communication patterns with the underlying hardware capabilities, ensuring smooth data flow across GPUs.
    \vspace{-1em}
\end{itemize}
From a system implementation perspective, existing systems such as Megatron~\cite{narayanan2021efficient} adopt a fully symmetric partitioning strategy, which often leads to GPU under-utilization due to the inherently asymmetric nature of heterogeneous environments. From a scheduling perspective, existing works---including Alpa~\cite{zheng2022alpa} and Galvatron~\cite{miao2022galvatron}---design high-complexity algorithms that cannot be trivially adapted to heterogeneous environments (detailed in \ref{sec:auto_parallel}). To overcome these challenges, we propose a novel framework, \sys, that coordinates distributed LLM training over a set of GPUs with different computation capabilities and connections. 

Our contributions are summarized as:

\noindent \textbf{\underline{Contribution 1.}} We implement \sys, a heterogeneous LLM training system, which supports asymmetric partition for the training computation within the scope of data, pipeline, and tensor model parallelism to flexibly accommodate the heterogeneity of various GPUs and diversified network connections. Such designs enable fine-grained partitioning of the original training computation to fully exploit the heterogeneous computational capabilities. 

\noindent\textbf{\underline{Contribution 2.}} We formulate the scheduling problem of allocating the LLM training computation over a set of heterogeneous GPUs as a constrained optimization problem. To solve this problem efficiently, we propose a two-phase optimization approach that employs a graph partitioning algorithm to effectively coordinate parallel strategies for the given devices. In the first phase, the algorithm splits available GPUs into multiple groups, each of which forms a pipeline in the second phase. We then iteratively apply the two-phase algorithm to determine optimal parallel strategies for the set of heterogeneous GPUs.

\noindent\textbf{\underline{Contribution 3.}} We evaluate \sys through experiments. We compare the system efficiency between heterogeneous settings (enabled by \sys) and standard homogeneous settings within a centralized data center (enabled by Megatron, Galvatron, and FSDP). We conduct these comparisons on training the popular LLM models with different model sizes, including \textsc{Llama-2 (7B)}, \textsc{Llama-2 (13B)}, and \textsc{Llama (30B)}. The results demonstrate that, given the \textit{same} theoretical \textsc{FLOPS}, \sys achieves performance comparable to state-of-the-art LLM training frameworks running in a homogeneous data center. \sys also operates efficiently in heterogeneous environments and outperforms state-of-the-art heterogeneous training systems (e.g., AMP~\cite{li2022amp}, Espresso~\cite{zhou2025espresso}, and Metis~\cite{um2024metis}). The results show that \sys offers $1.5\times$ to $2.4\times$ higher throughput. 

The rest of the paper is organized as follows. 
\S\ref{sec:preliminary} provides a brief review of optimization techniques for parallelizing LLM training and summarizes related work on auto-parallelization on homogeneous clusters and heterogeneity-aware training scheduling.
In \S\ref{sec:sys}, we present a case study and introduce the design of \sys, which utilizes fully asymmetric parallelism to optimize training in heterogeneous environments. 
In \S\ref{sec:scheduling}, we formalize the allocation of asymmetric partitioned training computations over a set of heterogeneous GPUs as a constrained optimization problem and propose an efficient solution based on a hierarchical graph partitioning algorithm. 
In \S \ref{sec:eval}, we evaluate the system performance of \sys and the scheduling algorithm.

\section{Preliminary and Related Work}
\label{sec:preliminary}

We first discuss representative work on automatic parallelization for homogeneous setups in \S\ref{sec:auto_parallel}, and finally summarize recent efforts on heterogeneity-aware training scheduling from the machine learning community in \S\ref{sec:rel}.

\vspace{-0.5em}
\subsection{Auto Parallelization on Homogeneous Clusters} 
\label{sec:auto_parallel}
Significant efforts have been made to automatically identify optimal parallel configurations for LLM training on homogeneous clusters~\cite{zheng2022alpa, miao2022galvatron, santhanam2021distir, chen2023autoddl, tarnawski2021piper, fan2021dapple}. For example, Alpa~\cite{zheng2022alpa} proposes a two-level hierarchical scheduling algorithm based on ILP and dynamic programming. 
Galvatron~\cite{miao2022galvatron} introduces another dynamic programming-based scheduling algorithm for transformer models.
These auto parallelization algorithms effectively identify optimal parallel configurations for homogeneous clusters. However, they assume symmetric setups in terms of network connectivity and hardware specifications, and employ high-complexity algorithms. Consequently, in heterogeneous environments, these algorithms often require prohibitively long execution times (e.g., several hours for hundreds of GPUs) while traversing a limited search space and producing parallel configurations that yield suboptimal performance. (See \S\ref{sec:eval}.)

\subsection{Heterogeneity-aware LLM Training} 
\label{sec:rel}

Significant efforts have been made to build training systems tailored for heterogeneous resources~\cite {miao2022galvatron,zheng2022alpa,unger2022unity,wang2024improving,yuan2022decentralized, yi2020optimizing, miao2021heterogeneity,zhang2022mics}. For example, SDPipe~\cite{miao2023sdpipe} implements flexible data parallelism synchronization schemes to address slow data parallel communication issues in (semi)-heterogeneous environments; Whale~\cite{jia2022whale} proposes a hardware-aware load-balancing algorithm to accelerate heterogeneous training. 

Among recent work, AMP~\cite{li2022amp}, Espresso~\cite{zhou2025espresso}, and Metis~\cite{um2024metis} represent the most relevant efforts toward supporting heterogeneous training and scheduling for the optimal parallel configurations (across data-, pipeline-, and tensor model-parallelism). AMP and Espresso are built on top of DeepSpeed~\cite{rajbhandari2020zero}; they support heterogeneous layer partitioning and perform scheduling within this design. However, they fall short of fully leveraging the potential of heterogeneous clusters due to limited support for heterogeneity-aware parallelism, which restricts the search space explored by their algorithms. Metis partitions computations into a single pipeline with a variable number of stages, each of which can have a distinct combination of data and tensor model parallelism degrees. To identify high-performance parallelism plans for heterogeneous GPUs, Metis introduces a novel depth-first-search (DFS) and heuristic-based approach (see \S4.2 of the Metis paper). Although Metis provides fine-grained heterogeneous-aware parallel support, it has two limitations: (\underline{i}) the introduced data-parallel-first DFS scheduler prioritizes data parallelism over tensor model parallelism unless an out-of-memory error occurs, whereas optimal throughput often comes from a balanced mix of data- and tensor model-parallelism. Due to these algorithmic limitations, strategy found by Metis's algorithm often only involve data and pipeline parallelism, resulting in inflated pipeline-bubbles, and up to 1.9$\times$ lower throughput than \sys; and (\underline{ii}) Metis's DFS-based search algorithm scales poorly: it runs for hours and still struggles to find optimal strategies for 240-GPU (see \S \ref{subsec:simulation}). In contrast, \sys is both capable of fully supporting the asymmetric partition of the parallel training computations (see \S \ref{sec:sys}, \S \ref{sec:eval}) and has a more efficient and effective scheduling algorithm in identifying the optimal parallel strategy (see \S \ref{sec:scheduling}). 

\vspace{-0.5em}
\section{System Design and Implementation}
\label{sec:sys}

In this section, we start with a case study on heterogeneous training with Megatron, then introduce the system design of \sys and discuss how \sys improves the training efficiency. We present full comparisons with homogeneous frameworks (Megatron), auto-parallelization work (Galvatron), and heterogeneous frameworks (AMP, Espresso, and Metis) in \S\ref{sec:eval}.

\subsection{Case Study: Parallelism over Heterogeneity}

Consider training a \textsc{Llama-2 (13B)} model in a heterogeneous environment on cloud with the following machines: machine \textit{A} is equipped with \texttt{3$\times$A800-80G} GPUs connected via NVLINK, which offers a intra-machine bandwidth of $200$ GB/s; machine \textit{B} has \texttt{3$\times$4090-24G} GPUs connected via PCIe with a bandwidth of $32$ GB/s; and machine \textit{C} is equipped with \texttt{2$\times$3090-24G} GPUs, connected via PCIe with a $16$ GB/s bandwidth.\footnote{3-GPU-host exists commonly in practice. For example, cloud providers (e.g., AWS~\cite{awsparallelcluster}) often manage resources via Slurm-based approaches, which may expose leftover 3-GPU when the rest five GPUs are allocated elsewhere.} The machines are interconnected using a $1$ GB/s Ethernet link. To motivate our system design compared to Megatron, we simulate performance on a training task with a global batch size of $24$, and a micro-batch size of $1$. Activation recomputation is applied for both systems. 

\begin{figure}
  \centering
  \includegraphics[width=\linewidth]{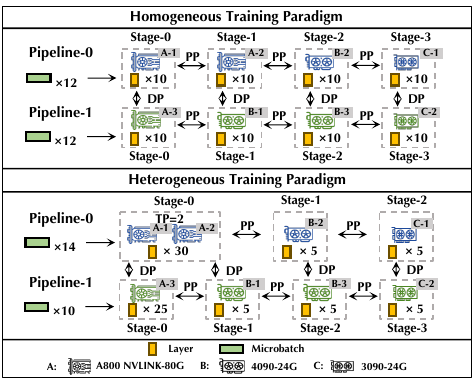}
  \caption{Case study on comparing Megatron and \sys. Both systems run their optimal parallel strategies.}
  \label{fig:example}
\end{figure}

\noindent \textbf{Training with Megatron.} Megatron only supports a fully symmetric partition of the training computation. Denote $D_{dp}$, $D_{tp}$, $D_{pp}$ as the parallel degrees for data-, tensor model-, and pipeline- parallelism. The considered parallel strategies on these $8$ GPUs include: (\underline{i}) plan 1: $(D_{dp}=1, D_{tp}=1, D_{pp}=8)$; (\underline{ii}) plan 2: $(D_{dp}=1, D_{tp}=2, D_{pp}=4)$; (\underline{iii}) plan 3: $(D_{dp}=2, D_{tp}=1, D_{pp}=4)$.

Plan 1 employs a pipeline with eight pipeline stages, which is inefficient due to high bubble cost and imbalanced computation. Even if the micro-batch size is set as 1 (which may reduce GPU kernel efficiency), bubble time is still over $22\%$ of the pipeline execution time. The system performance is further compromised by bottleneck machine \textit{C}, which computes 3.9$\times$ and 2.1$\times$ slower than machines \textit{A} and \textit{B}.  

Plan 2 introduces inter-machine tensor model parallelism, significantly degrading the performance with an estimated per-layer communication cost of $1.88$s, given $1$ GB/s bandwidth and batch size as $24$. This is a huge overhead, compared to the $0.01$s communication cost via NVLINK.

Plan 3 is a potentially good configuration. We fine-tune this plan by maximizing the number of transformer layers that use high intra-machine bandwidth for data parallel communication. As visualized in \autoref{fig:example}-(Top). This strategy creates four pipeline stages, each with 10 layers. However, even in this setting, Megatron has at least two deficiencies. First, two GPUs in machine \textit{A} run pipeline parallelism on pipeline-$1$, which wastes the high NVLINK intra-machine bandwidth and introduces higher pipeline bubble overhead. Second, GPUs in machine \textit{A} handle too few layers, wasting computation power.  

When the underlying heterogeneous environment mismatches with the fully symmetric strategy of Megatron, the system treats GPUs with strong computation capability as weak GPUs, and parallel strategies are constrained by the network connections, leading to the underutilization of computation resources. In this case, the state-of-the-art training system Megatron performs poorly, with an estimated end-to-end training iteration time of $41.52$s. 

\subsection{Asymmetric Parallel Support in \sys}
\label{sec:sys-degisn}

To train efficiently under heterogeneous settings, we implement \sys with fully asymmetric parallel support and system optimizations, with the essential change as: 

\noindent \textbf{Asymmetric partition of the computation load}. For each data-parallel replica\footnote{\ryan{For simplicity, in the remainder of the paper, we refer to each data-parallel replica as a pipeline.}}, which can be assigned a \textit{different} batch size, pipeline parallel communication groups are initialized with a \textit{different} tensor model parallel degree and a corresponding, potentially \textit{different} number of allocated transformer layers. Each pipeline stage selects a fixed leader GPU, which minimizes communication latency to GPUs in nearby stages and initializes a tensor model parallel group. During the forward pass, the leader GPU in each stage \texttt{send}s the activation to the leader GPU in the next stage. Once the leader GPU in the next stage \texttt{receive}s the activation, it \texttt{broadcast}s this activation among its tensor parallel group to perform computations. During the backward pass, the same logic applies for communicating the gradients w.r.t activations. \ryan{For cases where two adjacent pipeline stages have the same tensor model parallel degree, \sys falls back to the Megatron~\cite{narayanan2021efficient} scatter/gather communication strategy for collective-communication optimization.}

\noindent \textbf{Asymmetric gradient synchronization}. Model parameters (and corresponding gradients) of transformer layers across different pipelines may be chunked into different sizes due to differently assigned tensor parallel degrees. In this case, vanilla data parallelism cannot correctly synchronize gradients. To ensure communication correctness, we identify the smallest gradient size and further partition larger gradients into multiple chunks, each matching the size of the smallest identified gradient. Data parallel communication is then performed by synchronizing each gradient chunk among different subsets of data parallel workers, with the same communication volume.

\textbf{System implementation.} we optimize our system in four ways. First, we support gradient accumulation and activation recomputation to amortize data parallel communication cost and reduce memory footprint. Second, we leverage APIs in FlashAttention-2~\cite{daoflashattention}\footnote{\sys also supports FlashAttention-3~\cite{shah2024flashattention} on Hopper GPUs.} to create transformer layers with both tensor model parallelism and FlashAttention. We then implement our asymmetric pipeline parallelism---compatible with popular pipeline schedules such as GPipe~\cite{huang2019gpipe} and 1F1B~\cite{narayanan2019pipedream}---using these transformer layers. Third, we support asymmetric gradient synchronization logic by registering custom FSDP communication hooks. \ryan{Fourth, \sys incorporates a customized NCCL-based communication layer to support collective-communication across heterogeneous GPUs~\footnote{Note that only NVIDIA GPUs are supported in our implementation, which directly supports NCCL for different GPU types.}. \sys initializes NCCL-communication-groups for data-, tensor model-, and pipeline-parallelism to configure collective-communication across heterogeneous GPUs. GPUs within one communication group can communicate.}

\subsection{Case Study: Boost with \sys}

With the flexible design, \sys improves training efficiency in the former heterogeneous setting by creating two pipelines more efficiently as shown in \autoref{fig:example}-(Bottom). The improvements are as follows:

\noindent\textbf{Computation load is fully partitioned}. For each pipeline, the computation load is fully partitioned as follows. 

 \begin{itemize}[leftmargin=*]
 \vspace{-1em}
 
\item From the perspective of parallel strategy, we fully utilize the strong computational power and high intra-machine bandwidth in machine \textit{A} by applying tensor parallelism and assigning more transformer layers. Given the high intra-machine bandwidth in machine \textit{A}, tensor model parallelism is found superior to pipeline parallelism. Machine \textit{B} still applies pipeline parallelism as the intra-connection is not high enough for tensor model parallelism. 
\vspace{-0.5em}

\item From the perspective of layer partition, machine \textit{B} and machine \textit{C} undertake fewer transformer layers, which addresses the computation imbalance issue among pipeline stages. The number of layers that run inter-machine data parallelism is minimized. Megatron and \sys are both bounded by the slow data parallel communication on machine \textit{B}. Megatron communicates for $10$ transformer layers on machine \textit{B}, which can be estimated at $9.90$s. \sys only needs to communicate for $5$ transformer layers in $5.07$s, which is $1.9\times$ faster than Megatron.
\vspace{-0.5em}

\item From the perspective of pipeline efficiency, batch sizes are assigned differently to balance the computation speeds among pipelines. Otherwise, there will be no end-to-end improvement, as \sys remain bounded by the pipeline-$2$. The pipeline-$1$ has to wait for pipeline-$2$ to run data parallel communication. To unleash the efficiency of pipeline-$1$, we assign larger batch sizes to pipeline-$1$. In this way, we balance the execution time of each pipeline and improve the end-to-end performance. In our case study, two pipelines process batch sizes with a $40\%$ difference but exhibit a $7\%$ difference in execution time.
\end{itemize}

\noindent\noindent\textbf{Run asymmetric data parallelism}. For two GPUs in machine \textit{A} on stage-$0$ of pipeline-$1$, and GPUs on stage-$0$ and stage-$1$ of pipeline-$1$, although tensor parallelism partitions the parameters to different sizes, data parallel communication among these stages is supported as discussed in \S \ref{sec:sys-degisn}. This flexible design enhanced the performance of pipeline-$1$ from two aspects. (\underline{i}) Decreasing the number of pipeline stages reduces pipeline communication and bubble costs. (\underline{ii}) Adopting tensor parallelism on machine \textit{A} enhances the performance of pipeline-$1$. 

In summary, the end-to-end training iteration time of \sys for the parallel strategy shown in \autoref{fig:example}-(Bottom) is estimated to be $25.55$s, making it $1.6\times$ faster than Megatron in this hypothetical heterogeneous setting.

\vspace{-0.5em}
\section{Scheduling with Heterogeneity}
\label{sec:scheduling}

\textbf{Notations.} We summarize the notations used in the scheduling algorithm in \autoref{tab:notations}.

\begin{table}[htb]
\caption{Summarization of notations.}
\label{tab:notations}
\begin{center}
\begin{footnotesize}
\resizebox{\linewidth}{!}{
\begin{tabular}{c | l}
\toprule
\textbf{Symbol} & \textbf{Description} \\
\midrule
$d$  &  GPU device. \\
$m_d$      & GPU memory of device $d$. \\
$c_d$      & Tensor core computation power of device $d$. \\
$\alpha_{d,d'}$  & Latency between devices $d$ and $d'$. \\
$\beta_{d,d'}$   & Bandwidth between devices $d$ and $d'$. \\
$\sigma$   & Parallel execution plan of devices in $\mathbf{D}$. \\
$D_{dp}$   & Number of pipelines. \\
$\mathbf{D}$  &  Set of $N$ GPU devices $d_1, d_2, ..., d_N$. \\
$\mathbf{E}$   & Edge set of global graph. \\
$G=(\mathbf{D}, \mathbf{E})$   & Global graph for GPU set $\mathbf{D}$. \\
$\mathbf{P}$  & Global partition containing GPU set $\mathbf{D}_1, ..., \mathbf{D}_{D_{dp}}$. \\
$k_i$ & Number of GPU groups generated in $i$-th pipeline. \\
$\mathbf{D}_i$   & GPU set of the $i$-th pipeline. \\
$\mathbf{E}_i$   & Edge set of the $i$-th pipeline. \\
$G_i=(\mathbf{D}_i, \mathbf{E}_i)$   & Secondary graph of GPU set $\mathbf{D}_i$. \\
\multirow{2}{*}{$\mathbf{P}_i$}   & Secondary partition containing GPU set \\
   & $\mathbf{D}_{i,1}, ..., \mathbf{D}_{i, k_i}$. \\
$\mathbf{D}_{i,j}$   & GPU set of the $i$-th pipeline, $j$-th GPU group. \\
$\tau$   & Parameter for searching pipeline stage order. \\

\bottomrule
\end{tabular}
}
\end{footnotesize}
\end{center}

\end{table}


\begin{figure*}
  \centering\includegraphics[width=\linewidth]{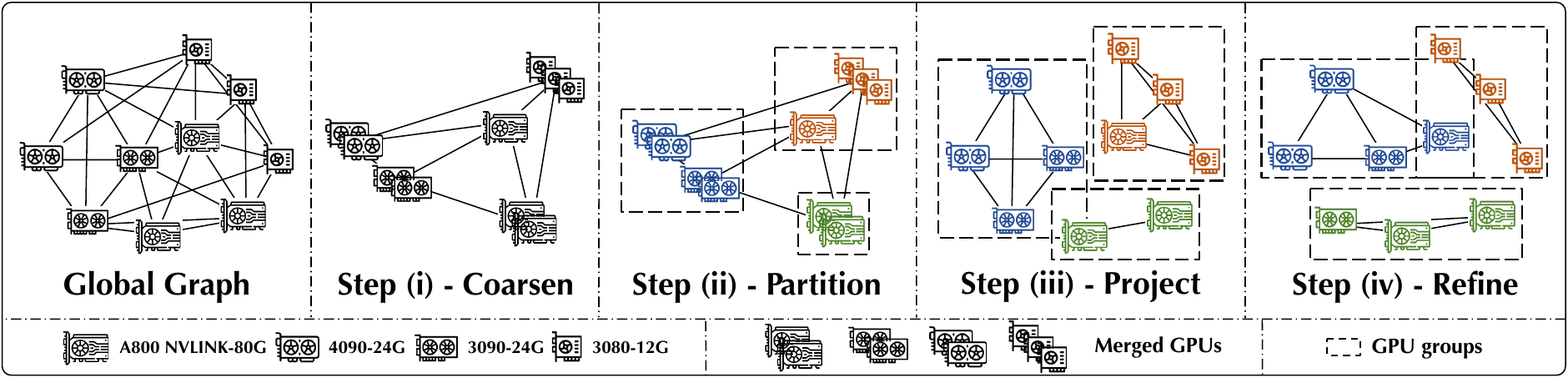}
  \caption{First phase: the global graph is partitioned into three groups of GPUs by four steps: (\underline{i})-coarsen, (\underline{ii})-partition, (\underline{iii})-project, and (\underline{iv})-refine. GPUs in the global graph are divided into three groups, which will be constructed as three pipelines.}
  \label{fig:global_partitioning}
\end{figure*}

\subsection{Formalization of the Scheduling Problem}
\label{formalize}

Given a set of heterogeneous GPUs, we aim to identify the optimal heterogeneous parallel execution strategy that minimizes the training iteration time. We formalize the scheduling problem as follows. Let $\mathbf{D} = \{d_1 \ldots d_N \}$ be a set of $N$ GPU devices, and the GPU device memory limit be notated as $m_d$. Given a particular GPU set, the scheduling problem can be defined as identifying the optimal parallel execution plan $\sigma^*$ that minimizes the training iteration execution time under the constraint of memory consumption:
\vspace{-2em}

\begin{small}
\begin{equation}
\begin{aligned}
\sigma^{*} = \arg \min_{\sigma} \quad & \textsc{Comm-Cost}\left( \sigma \right) + \textsc{Comp-Cost}\left( \sigma \right) \\
s.t.  \quad & \textsc{Mem-Cumsum}\left( d \right) \leq m_d \quad \forall d \in \mathbf{D} 
\end{aligned}
\end{equation}
\end{small}%

\vspace{-1em}
\noindent where $\textsc{Comm-Cost}\left( \sigma \right)$ and $\textsc{Comp-Cost}\left( \sigma \right)$ represent communication and computation time costs (we present the detailed cost modeling in Appendix \ref{app:cost_model}). of one iteration given a parallel execution plan $\sigma$. $\textsc{Mem-Cumsum}\left( d \right)$ represents the memory consumption of device $d$.

One parallel execution plan $\sigma$ can involve an arbitrary number of pipelines, each with varying global batch sizes, micro-batch sizes, and parallel strategies. Furthermore, a particular parallel strategy can have different configurations for data, pipeline, and tensor model parallel degrees. Each pipeline stage can contain a flexible number of transformer layers. Finding the exact optimal parallel strategy—considering the computation costs, communication costs, and memory consumption of all potential configurations—is NP-hard due to the exponential scale of candidate allocations.

Therefore, in a heterogeneous cluster with varying GPU capacities and network connectivity, it is often only possible to identify a near-optimal parallel execution plan using heuristic-based scheduling algorithms. Naively adopting the dynamic programming-based algorithms proposed in Alpa~\cite{zheng2022alpa} or Galvatron~\cite{miao2022galvatron} leads to poor scheduling results or requires a significantly longer search time. To address the challenge in efficiently identifying a near-optimal parallel execution plan in heterogeneous clusters, we design a two-phase scheduling algorithm to find a parallel execution plan $\sigma$ and iteratively optimize to a near-optimal solution $\sigma^*$. To be more concrete:
\begin{itemize}[leftmargin=*]
\vspace{-1em}
\item We introduce the first phase algorithm in \S \ref{sec:first-phase} that partitions the device set $\mathbf{D}$ into multiple GPU groups, each of which will be used to create one pipeline;
\vspace{-0.5em}

\item We enumerate the second phase algorithm in \S \ref{sec:second-phase} that identifies parallel execution plan $\sigma$ for each pipeline;
\vspace{-0.5em}
\item We iteratively repeat the two-phase algorithm and optimize the parallel execution plan to $\sigma^*$, illustrated in \S \ref{algo:iterate}.
\vspace{-0.5em}

\end{itemize}

\subsection{First Phase of Scheduling Algorithm}
\label{sec:first-phase}
The key insight of the first phase algorithm is to partition the GPUs into multiple groups, with each group forming a separate pipeline. Data parallel communication is then performed among these pipelines to synchronize gradients.

Assume we divide devices set $\mathbf{D}$ into $D_{dp}$ groups of GPUs in the current iteration, and \textit{minimize} bandwidth for data parallel communication (we discuss parameter $D_{dp}$ and bandwidth allocation in \S \ref{algo:iterate}). We first organize all GPUs from devices set $\mathbf{D}$ as a global graph $G=(\mathbf{D}, \mathbf{E})$, where each GPU $d \in \mathbf{D}$ is a vertex, and computation power $c_d$ is vertex weight; $\forall d_1, d_2 \in \mathbf{D}$, communication bandwidth $\beta_{d_1, d_2}$ represent an edge between these two GPUs in set $\mathbf{E}$. Then we partition global graph $G$ into global partition $\mathbf{P} = \{\mathbf{D}_1, \mathbf{D}_2, ..., \mathbf{D}_{D_{dp}} \}$, where each set $\mathbf{D}_i \in \mathbf{P}$ contains GPUs with high bandwidth that are used in $i$-th pipeline , and $\mathbf{D}_i \cap \mathbf{D}_j=\emptyset, \forall i,j$.  We adopt a partitioning method based on a $D_{dp}$-way multi-level graph partition algorithm~\cite{hendrickson1995multi} consisting of four steps:

\noindent {\textbf{Step (\underline{i}) - Coarsen}}. The global graph is coarsened into smaller graphs to simplify graph partition. Directly partitioning a large global graph (i.e., including many vertices) into $D_{dp}$ parts is usually inefficient. In contrast, partitioning a smaller coarsened graph is more efficient. We adopt \textit{heavy edge matching (HEM)} algorithm~\cite{karypis1998fast} for coarsening. To allocate low bandwidth for data parallelism, this coarsening operation merges GPUs with \textit{high} bandwidth. \ryan{The bandwidth connecting GPUs within super-nodes is considered high when it is relatively higher than the bandwidth connecting GPUs from different super-nodes.} As illustrated in \autoref{fig:global_partitioning}, step (\underline{i}), the coarsened graph contains only half as many vertices as the global graph.

\noindent  {\textbf{Step (\underline{ii}) - Partition}}. The coarsened graph is partitioned into $D_{dp}$ GPU groups, ensuring network bandwidth among GPU groups is minimized. With a recursive bisection method~\cite{karypis1998multilevel}, we recursively bisect the coarsened graph until $D_{dp}$ parts partition is obtained. Graph partition in this step solves a constrained partition problem that \textit{minimizes} $\textsc{Cut}$ objective function~\cite{wei1989towards} under two constraints: (\underline{i}) maintaining strict balance, and (\underline{ii}) partitioning to exactly $D_{dp}$ parts. $\textsc{Cut}$ objective function is defined by two levels: (\underline{i}) $\textsc{Cut}$ function between any two sets $\mathbf{D}_i, \mathbf{D}_j \in \mathbf{P}, \forall i,j$ is defined as the sum of edge weights (i.e., bandwidths) that connect them. (\underline{ii}) $\textsc{Cut}$ objective function for global partition $\mathbf{P}$ is defined as the summation of all cuts between sets $\mathbf{D}_i, \mathbf{D}_j \in \mathbf{P}, \forall i,j$. Formally, the two-level $\textsc{Cut}$ function is defined as:
\vspace{-1em}

\begin{small}
\begin{equation}
\label{eq:2}
\begin{aligned}
 \textsc{Cut}(\mathbf{D}_i, \mathbf{D}_j) = & \sum_{d_i \in \mathbf{D}_i} \sum_{d_j \in \mathbf{D}_j} \beta_{d_i,d_j}, \quad \forall \mathbf{D}_i, \mathbf{D}_j \in \mathbf{P} \\
 \textsc{Cut}(\mathbf{P}) = & \sum_{\mathbf{D}_i, \mathbf{D}_j \in \mathbf{P}} \textsc{Cut}(\mathbf{D}_i, \mathbf{D}_j) 
\end{aligned}
\end{equation}
\end{small}%

\vspace{-1em}
The constraint that measures the balance of global partition $\mathbf{P} = \{\mathbf{D}_1, \mathbf{D}_2, ..., \mathbf{D}_{D_{dp}} \}$ is defined as the maximum sum of vertex weights $\max_{\mathbf{D}_i \in \mathbf{P}} \sum_{d \in \mathbf{D}_i} c_d$, over the average sum of vertex weights $\frac{\sum_{d \in \mathbf{D}} c_d}{D_{dp}}$. This balance factor is always greater than or equal to $1$; its maximum value is a hyperparameter. A value closer to $1$ indicates more evenly distributed total vertex weights among GPU sets $\mathbf{D}_i$.

\noindent \textbf{{Step (\underline{iii}) - Project \& Step (\underline{iv}) - Refine}}. Partitioning coarsened graph is not the ultimate goal---as illustrated in \autoref{fig:global_partitioning}, step (\underline{iii}), to find the partition of global graph $G$, we project the results back, i.e., apply reverse operation of step (\underline{i}) to recover the $D_{dp}$-parts of the partition in global graph $G$. To effectively consider information within coarsened nodes, a refinement algorithm is necessary to enhance partition quality and maintain balance; for this purpose, we employ the Kernighan-Lin algorithm~\cite{kernighan1970efficient} in step (\underline{iv}) to adjust partition results.

\begin{figure}[t]
  \centering
  \includegraphics[width=\linewidth]{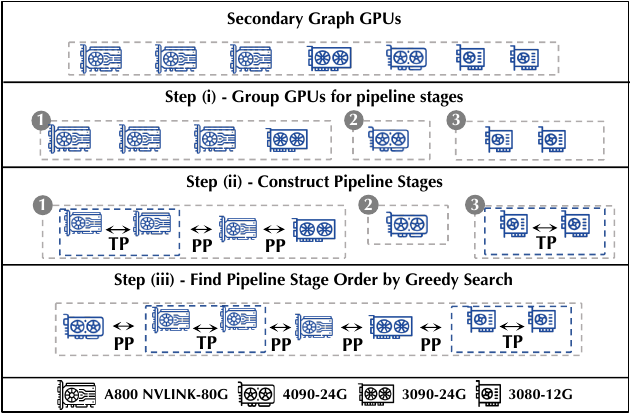}

  \caption{Second phase: each pipeline is created in three steps. (\underline{i}) GPUs with high bandwidth connections are grouped by graph partition. (\underline{ii}) \textit{intra-group strategy} is searched separately for each machine, i.e., GPUs in the same machine. (\underline{iii}) Pipeline stage order is determined by permuting all intra-group strategies by a top-$\tau$ greedy search algorithm.}
  \label{fig:secondary_partitioning}
\end{figure}

\subsection{Second Phase of Scheduling Algorithm}
\label{sec:second-phase}
The key insight of this phase is to efficiently generate the pipeline layout based on the graph partition results from the first phase. Thanks to the flexible asymmetric data parallelism design, we can independently determine a parallel strategy for each pipeline. However, the search space remains large, as we must determine the pipeline and tensor model parallelism strategy, as well as the execution order of the pipeline stages. In a fully heterogeneous environment, carefully permuting pipeline stages is necessary due to the heterogeneity of network connections. Formally, given the $D_{dp}$ groups of GPUs, we use GPUs in each GPU set $\mathbf{D}_i \in \mathbf{P}$ to find a parallel execution plan $\sigma_i$ for each pipeline. Finding the near-optimal layout for the assigned GPUs involves three key steps: (\underline{i}) grouping GPUs for pipeline stages based on graph partition; (\underline{ii}) constructing pipeline stages within each GPU group; and (\underline{iii}) determining the order of pipeline stages under a heterogeneous network.

\noindent \textbf{Step (\underline{i}) - Group GPUs for pipeline stages}. To group GPUs with high bandwidth connections and introduce algorithmic convenience to determine the stage order of the $i$-th pipeline (introduced shortly), we group GPUs with high bandwidth connections by further splitting GPUs in set $\mathbf{D}_i \in \mathbf{P}$ into multiple groups. Concretely, we first organize each set $\mathbf{D}_i \in \mathbf{P}$ into a secondary graph $G_i=(\mathbf{D}_i, \mathbf{E}_i), i=1,..., D_{dp}$, where the edge set $\mathbf{E}_i$ contains communication bandwidths connecting GPUs in set $\mathbf{D}_i$. Next, we partition each secondary graph $G_i$ into secondary partition $\mathbf{P}_i=\{ \mathbf{D}_{i,1}, ..., \mathbf{D}_{i, k_i} \}$, where each parameter $k_i$ controls the number of parts that $G_i$ is partitioned into. Set $\mathbf{D}_{i,k} \in \mathbf{P}_i$, contains GPUs to construct pipeline stages, and $\mathbf{D}_{i,k_1} \cap \mathbf{D}_{i,k_2}=\emptyset, \forall k_1, k_2$. The secondary graphs are partitioned using the same multi-level graph partition method as discussed in \S \ref{sec:first-phase}. As an illustrative example shown in \autoref{fig:secondary_partitioning}-(Top), GPUs in a secondary graph are first partitioned into three groups after this step.
Note that GPUs within the same GPU sets $\mathbf{D}_{i,k} \in \mathbf{P}_i$ have high bandwidth connections. Finding the stage order among these stages has minor effects since pipeline communication costs are not the bottleneck. In contrast, when pipeline stages are generated by different GPU sets $\mathbf{D}_{i,k} \in \mathbf{P}_i$, permuting these stages can significantly reduce the pipeline communication overhead by effectively utilizing the low bandwidth. 

In the following two steps, we first construct pipeline stages within each GPU set $\mathbf{D}_{i,k} $ without considering pipeline stage order. Then, we search the pipeline stage order among pipeline stages created by different GPU sets.

\noindent \textbf{Step (\underline{ii}) - Construct pipeline stages}. Given the secondary graph partition results of each secondary graph $G_i$, we construct the $i$-th pipeline as follows. Each GPU set $\mathbf{D}_{i,k} \in \mathbf{P}_i$ separately searches its \textit{intra-group strategy} within each machine, through simulating different parallelism strategies with cost models defined in the Appendix \ref{app:cost_model} and selecting the locally optimal parallel strategy. As shown in \autoref{fig:secondary_partitioning}-(Middle), the first GPU group constructs three stages, while other GPU groups construct one stage in each.

\noindent \textbf{Step (\underline{iii}) - Find pipeline stage order by greedy search}. 
We consider each \textit{intra-group strategy} as a single vertex, and construct a new graph $G'_i$ for the $i$-th pipeline. The stage order for the $i$-th pipeline is thereby searched by a top-$\tau$ greedy algorithm. The algorithm runs in two nested loops. First, it selects each GPU group as the starting group. Second, for each neighboring GPU group with $\tau$-highest inter-group bandwidth, the algorithm recursively explores their neighboring GPU groups with $\tau$-highest inter-group bandwidth until a pipeline path is generated. As shown in \autoref{fig:secondary_partitioning}-(Bottom), stage order within each \textit{intra-group strategy} is not changed (as we do not permute them), while three pipeline stages in the first \textit{intra-group strategy} are placed after the one pipeline stage in the second \textit{intra-group strategy}, and before the one pipeline stage in the last \textit{intra-group strategy}.

\subsection{Iterative Optimization}
\label{algo:iterate}

Lastly, we introduce the iterative optimization procedure --- The parallel execution plan $\sigma$ is iteratively optimized to the final near-optima $\sigma^*$ from two aspects:

\noindent \textbf{Optimize from first phase algorithm:} The first phase algorithm can be optimized from two aspects. 

First, the algorithm partitions the global graph into different numbers of pipelines by enumerating the parameter $D_{dp}$ across iterations to optimize the number of pipelines. 
\textcolor{red}{
}

Second, the algorithm carefully allocates network bandwidth for data and pipeline parallelism communication. One of our key observations is that both data parallel communication and pipeline execution time can be the bottleneck in heterogeneous environments. When pipelines have many pipeline stages and handle a large batch size, pipeline execution time accounts for most of the training iteration time. In this case, minimizing bandwidth for data parallelism and maximizing bandwidth for pipeline execution can effectively improve the system performance. The system performance benefits from the reduced communication overhead of pipeline execution. On the other hand, when pipelines have few pipeline stages and handle a small batch size, system performance benefits from maximizing bandwidth for data parallelism and minimizing bandwidth for pipeline execution. The system performance is boosted by the reduced data parallel communication overhead. Based on this observation, we implement two partition options: either (i) \textbf{maximize} or (ii) \textbf{minimize} the inter-group (i.e., GPU groups for each pipeline) edge weights (i.e., bandwidth). Maximizing the inter-group edge weights corresponds to maximizing the \textsc{Cut} objective function in \autoref{eq:2}, which in turn results in allocating high communication bandwidth for data parallelism; conversely, minimizing the edge weights results in allocating low communication bandwidth for data parallelism. At each iteration, we adaptively select the potentially optimal partition option based on the historical moving average of costs to optimize system performance. Specifically, we simulate data parallel communication and pipeline execution costs at the end of each iteration and update the historical average costs. In the next iteration, the partition decision is made using this historical information. With this design, our first-phase algorithm effectively allocates network bandwidth, enhancing system performance.

\noindent \textbf{Optimize from second phase algorithm:} Given the global graph partition result in the first phase, varying the parameters $k_i$ across iterations results in a distinct intra-group strategy, which, in turn, determines the configuration of pipeline stages and their order. By fine-tuning $k_i$, we can construct pipelines that achieve high efficiency.

 \ryan{\noindent \textbf{Batch and layer assignment:} For a given parallel execution plan generated by our two-phase algorithm, our algorithm automates batch and layer distribution using \textit{greedy strategies}. For batch assignment, we distribute batches proportionally to the estimated training efficiency of each pipeline. For layer assignment, we prioritize assigning layers to pipeline stages in proportion to their estimated computational efficiency. If out-of-memory issues are detected on certain stages, we reduce the number of layers assigned to those stages and reassign the corresponding layers to stages with the most abundant memory resources.
}

\noindent \textbf{Evaluate by simulation:} at the end of each iteration, we simulate the execution costs for the generated parallel execution plan using our cost model, given the communication bandwidth matrix and hardware computation capabilities. The communication bandwidth matrix serves as an abstraction of the entire cloud network, and we obtain this matrix through profiling communication status between every GPU-pair; for computation capabilities, we represent them by theoretical peak FLOPS of each GPU type. When a generated parallel execution plan results in an out-of-memory error, its execution cost is set to infinity. If the plan results in poor performance due to imbalanced computation, inefficient memory usage, or suboptimal network utilization, the cost is assigned a high penalty value. Conversely, if the plan efficiently utilizes the given heterogeneous cluster, a low cost is assigned. Finally, the parallel strategy with the lowest estimated cost is selected.
Our cost model closely predicts actual execution costs. We provide a detailed evaluation on the simulation accuracy in \S\ref{subsec:algo} and Appendix \ref{app:cost_model_val}, results show that simulation deviations are less than $2.9\%$. To enhance simulation accuracy, we further incorporate network latency ($\alpha_{d,d'}$) into the simulation. As the number of collective communication operations increases, system performance degrades due to high linking costs. Network latency is particularly critical in heterogeneous environments, where a large number of micro-batches and long pipeline schedules lead to an increased number of NCCL operations.

\section{Evaluation}
\label{sec:eval}



\subsection{Experimental Setups}
\vspace{-0.5em}
\noindent \textbf{Experimental setup.} To thoroughly compare the end-to-end performance of \sys and state-of-the-art frameworks, we include Megatron, Galvatron, and FSDP as baseline frameworks, and \textsc{Llama} models in different scales as representative models\footnote{LLM often differ on model scales, instead of model structure.}. We evaluate \sys based on the following experiment settings:

\textit{Homogeneous settings.} We rent two \texttt{8$\times$A800 PCIe-80G} with or without RDMA, to test the maximum throughput (measured by the achieved PFLOPS) on \textsc{Llama-2 (7B)}, \textsc{Llama-2 (13B)}, and four \texttt{8$\times$A800 PCIe-80G} to test the maximum throughput on \textsc{Llama (30B)}. 

\textit{{Heterogeneous settings.}} We rent GPUs from cloud provider \texttt{Ucloud}. To evaluate system performance in various heterogeneous environments, we test and compare the performance of Megatron, Galvatron, and \sys under three settings (inter-connections at $5.6$ Gbps) as follows:

\begin{itemize}[leftmargin=*]
\vspace{-1em}
\item \textit{Heterogeneous setting 1}: one \texttt{8$\times$3080Ti}, one \texttt{8$\times$3090}, and three \texttt{8$\times$4090}. With $1.36\%$ higher total \textsc{FLOPS} and $1.48\times$ smaller total memory, we compare the maximum throughput gap with two \texttt{8$\times$A800 PCIe-80G} in training \textsc{Llama-2 (7B)/(13B)}. \texttt{3080Ti} has 12 GB memory, \texttt{3090} and \texttt{4090} each have 24 GB memory. Intra-machine connections of \texttt{3080Ti} and \texttt{3090} are $24$ GB/s.
\vspace{-0.5em}
\item \textit{Heterogeneous setting 2}: one \texttt{8$\times$3080Ti}, one \texttt{8$\times$3090}, one \texttt{8$\times$4090}, and one \texttt{8$\times$A800 NVLINK-80G} (200 GB/s NVLINK). With 1.59\% less total theoretical \textsc{FLOPS} and $1.14\times$ smaller total memory, we compare the maximum throughput gap with two \texttt{8$\times$A800 PCIe-80G} in training \textsc{Llama-2 (7B)/(13B)}.
\vspace{-0.5em}
\item \textit{Heterogeneous setting 3}: one \texttt{8$\times$3090}, two \texttt{4$\times$3090},
four \texttt{8$\times$4090}, and one \texttt{8$\times$A800 NVLINK-80G}. With 4.67\% less total theoretical \textsc{FLOPS} and $1.43\times$ smaller memory, we compare the maximum throughput gap with four \texttt{8$\times$A800 PCIe-80G} in training \textsc{Llama (30B)}.
\vspace{-0.5em}
\end{itemize}

\subsection{End-to-end Performance}
\label{subsec:e2e}

\begin{figure}
    \centering
    \includegraphics[width=\linewidth]{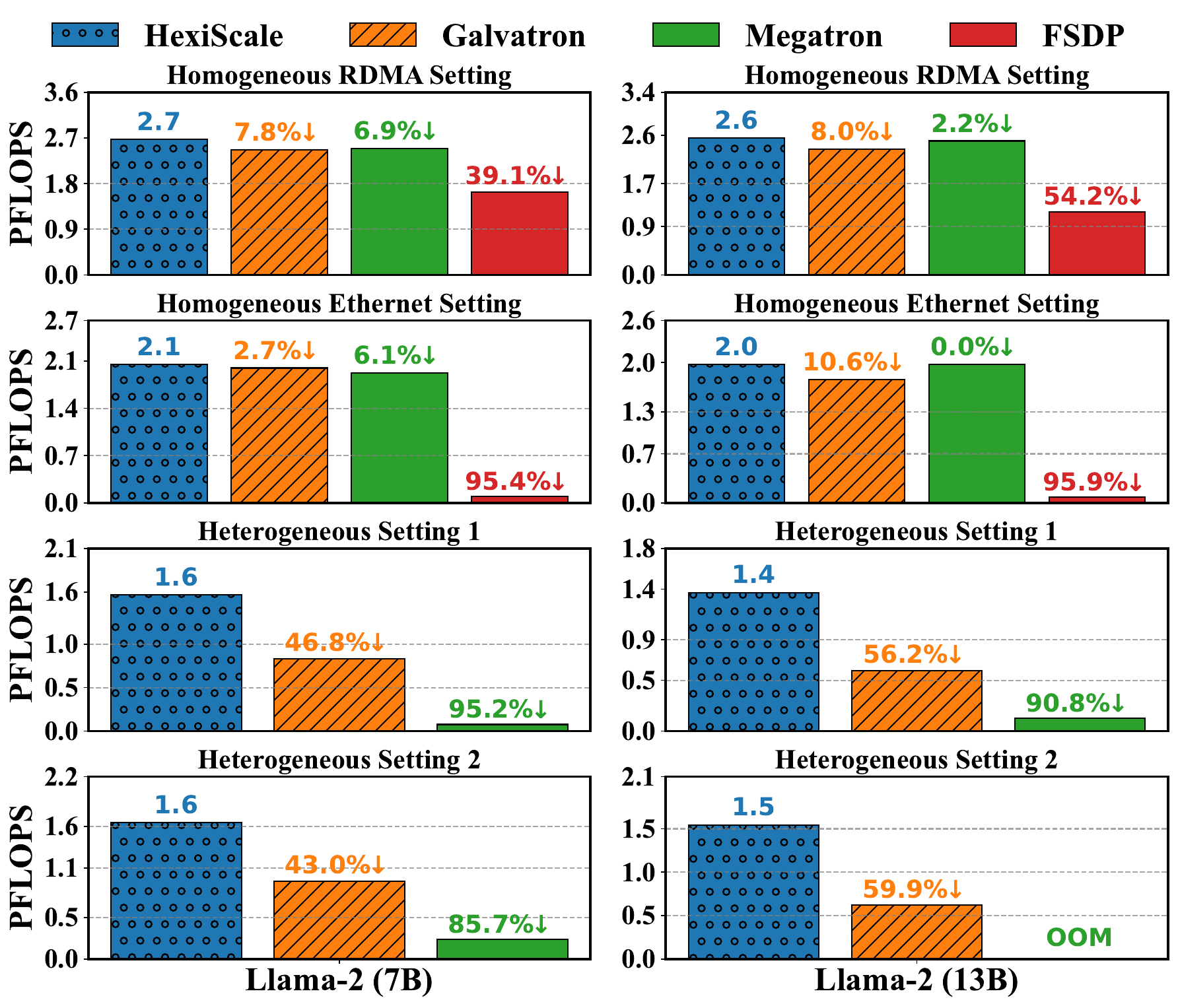}
    \vspace{-1em}
    \caption{End-to-end experiments of \sys compared with other systems under various settings on \textsc{Llama-2 (7B / 13B)}.}
    \label{fig:e2e}
    \vspace{-1em}
\end{figure}

\noindent \textbf{Results and discussions.} \noindent\textit{First, \sys exhibits performance comparable to other high-performance systems in homogeneous settings.} As shown in \autoref{fig:e2e} and \autoref{fig:e2e-30b}~\footnote{\ryan{We present the detailed parallel execution configurations in Appendix \ref{app:eval_detail} for each system.}}, \sys\ achieves comparable performance as the baseline frameworks Megatron and Galvatron under both RDMA and Ethernet inter-machine connections. FSDP is unsuitable for clusters with low-speed inter-connections; even with RDMA bandwidths of $10$ GB/s, using FSDP alone exhibits poor performance.

\textit{Second, comparisons across frameworks in heterogeneous settings highlight the adaptability of \sys.} While Megatron and Galvatron perform well in homogeneous environments, they cannot be easily adapted to heterogeneous settings. As shown in \autoref{fig:e2e-30b}, Megatron can not run in heterogeneous setting 3 when training \textsc{Llama (30B)}. Despite tuning various parallel strategies, out-of-memory issues persist. Megatron enforces a fully symmetric parallel strategy, leading to significant imbalance issues. Different types of GPUs cannot fully utilize computational capabilities and intra-machine bandwidth when restricted to a uniform parallel strategy. For example, GPUs with high intra-machine bandwidth must adopt a low tensor model parallel degree to accommodate GPUs with lower bandwidth, resulting in underutilized resources. Galvatron offers greater flexibility than Megatron by allowing flexible transformer layer assignment. However, parallel strategies remain suboptimal for heterogeneous clusters, leading to performance degradation. Compared to Galvatron, \sys achieves up to a $2.5\times$ higher throughput and, on average, a $2.1\times$ improvement.

\begin{figure}
    \centering
    \includegraphics[width=\linewidth]{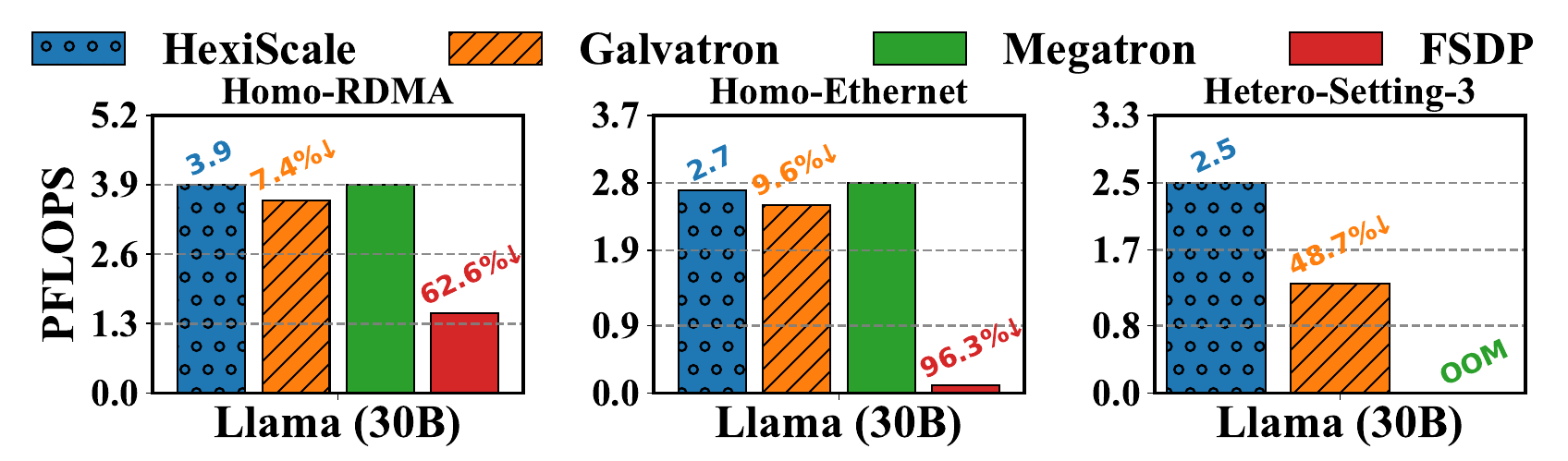}
    \vspace{-1em}
    \caption{End-to-end experiments of \sys{} compared with other systems under various settings on \textsc{Llama (30B)} model.}
    \vspace{-1em}
    
    \label{fig:e2e-30b}
\end{figure}
\textit{Finally, comparisons between homogeneous settings and heterogeneous settings demonstrate the strong competitiveness of \sys.} In homogeneous settings with RDMA, the achieved maximum throughput of Megatron and Galvatron is significantly higher than when training with \sys in heterogeneous environments. This performance gap arises from inter-machine connections: Megatron and Galvatron utilize high-performance RDMA, whereas \sys runs over slower Ethernet-based inter-connections. The performance of \sys can be further improved by replacing Ethernet with RDMA connections (This point is validated through simulation, as presented in \S \ref{subsec:simulation}). To fairly evaluate heterogeneous training performance, we conduct baseline experiments using Megatron, Galvatron, and FSDP in homogeneous settings with Ethernet inter-machine connections. As shown in \autoref{fig:e2e} and \autoref{fig:e2e-30b}, despite the challenges posed by imbalanced GPU computation, limited total memory capacity, and more frequent inter-machine communication in heterogeneous clusters, \sys exhibits up to $1.01\times$ throughput and an average throughput gap of $0.83\times$ compared to homogeneous scenarios~\footnote{\ryan{The throughput gaps are calculated by dividing the throughput of \sys on heterogeneous clusters by the throughput of Megatron and Galvatron on homogeneous clusters interconnected via Ethernet.}}. By asymmetric system design and effective scheduling, \sys better utilizes fragmented GPU resources and exhibits strong potential in diverse heterogeneous settings.

\subsection{Ablation Studies}
\label{subsec:ablation}

\begin{figure}
    \centering
    \includegraphics[width=\linewidth]{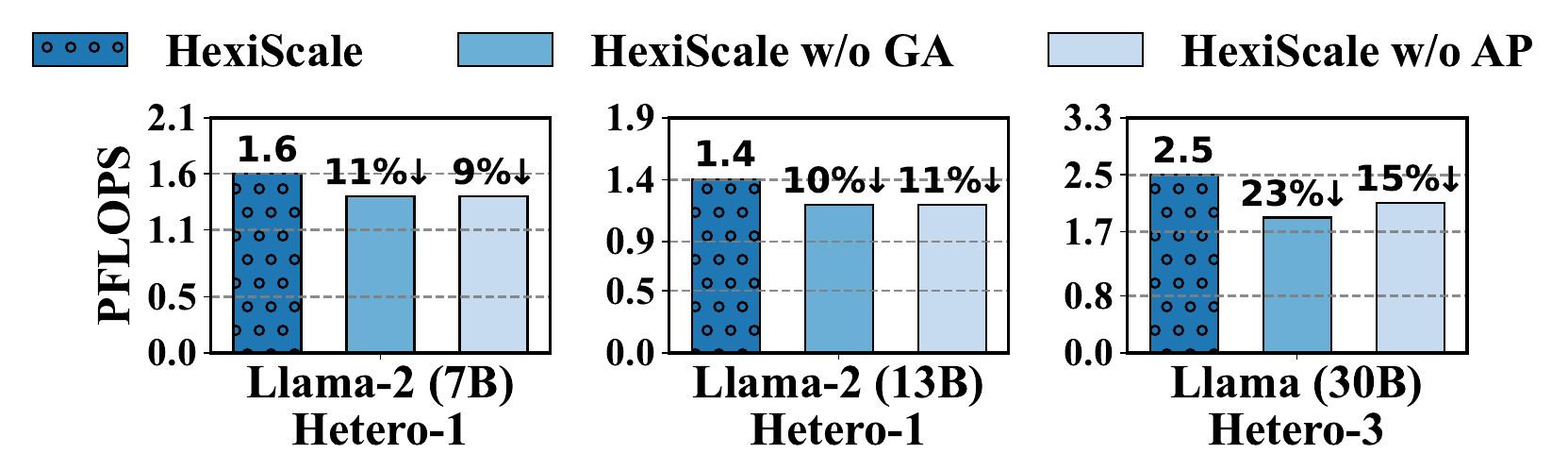}
    \vspace{-1em}
    \caption{Breakdown experiments of \sys with \textsc{Llama-2 (7B / 13B / 30B)} models under hetero-setting 1 and 3.}
    \label{fig:ab}
    \vspace{-1em}
\end{figure}

\noindent\textbf{System design breakdown.} \sys is implemented with asymmetric pipeline parallelism, under the support of asymmetric data parallelism, and other system optimizations, including gradient accumulation (GA). We evaluate the effectiveness of each design separately, the results are shown in \autoref{fig:ab}. We analyze the ablation results as follows:

\textit{Disabling our asymmetric parallel support}, system performance at most degrades $23\%$ (in training \textsc{Llama (30B)}), and $15\%$ on average. Without our asymmetric parallel support, all pipeline stages must have the same tensor model parallelism degrees, which is often suboptimal due to the underutilization of distinct hardware features. For example, in heterogeneous setting 3, a higher tensor model parallelism degree on \texttt{8$\times$A800} is beneficial. Conversely, GPUs with lower intra-machine bandwidths should be assigned lower tensor model parallelism degrees (detailed in \S \ref{subsec:algo}).

\textit{Disabling gradient accumulation}, system performance at most degrades $15\%$ (for \textsc{Llama (30B)}), and $12\%$ on average. Heterogeneous GPUs with limited memory can only accommodate a small batch size. As a result, the execution time between two data parallel gradient synchronizations is short. Without gradient accumulation, frequent data parallel communication degrades system performance.


\begin{figure}
    \centering
    \includegraphics[width=\linewidth]{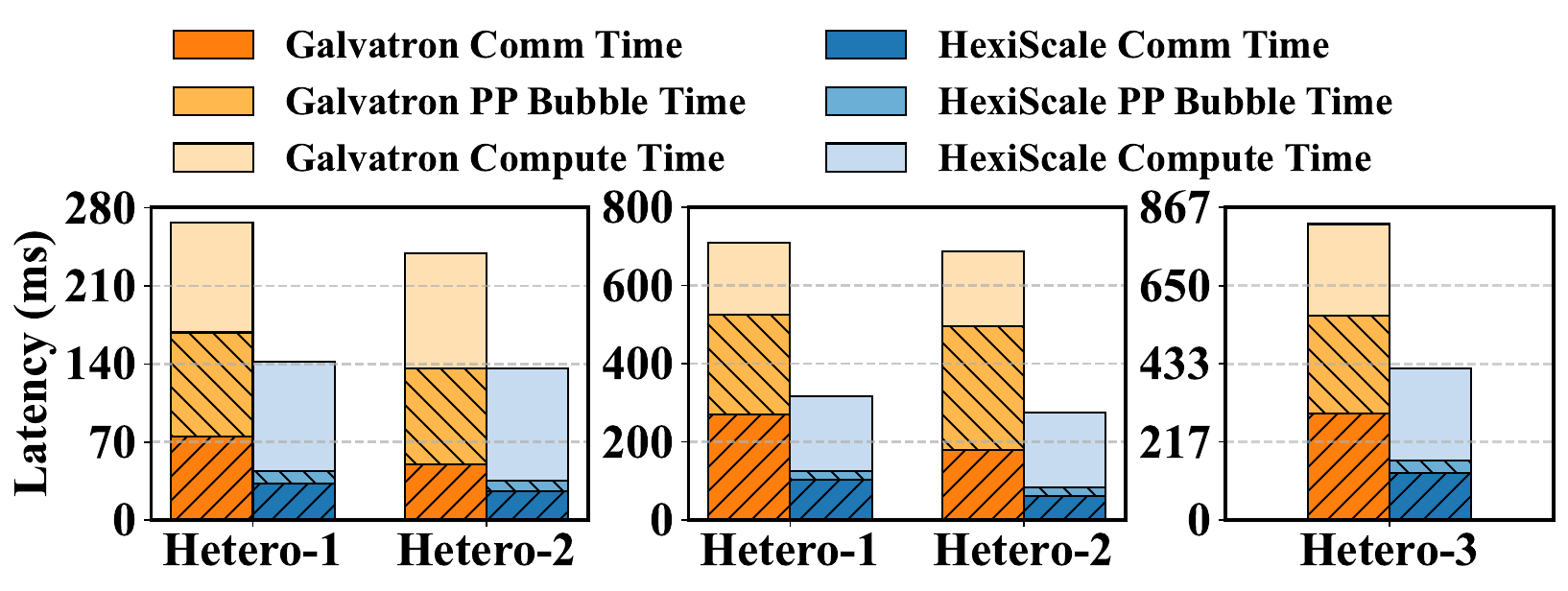}
    \vspace{-1em}
    \caption{Breakdown of end-to-end time across different heterogeneous experimental settings and models. We benchmark the per-batch communication time, computation time, and pipeline bubble time for \sys and Galvatron.}
    \label{fig:lat_bk}
    \vspace{-1em}
\end{figure}

\noindent
\textbf{Training iteration latency breakdown.} We present breakdown of training iteration latency of \sys and Galvatron under various heterogeneous settings in~\autoref{fig:lat_bk}. \sys outperforms Galvatron by reducing communication overhead and pipeline bubbles. With a symmetric parallel strategy, Galvatron runs a high degree of pipeline parallelism, which increases both communication overhead and pipeline bubbles. Furthermore, Galvatron experiences additional bubbles due to imbalanced computation across pipeline stages. With asymmetric parallel support, \sys addresses these challenges by reducing the number of pipeline stages and carefully balancing computation.

\subsection{Scheduling Algorithm Evaluation}
\label{subsec:algo}
Another major concern is the performance of our algorithm, including the scheduling results, simulation accuracy, effectiveness of graph partitioning, and running efficiency.

\begin{figure}[t]
  \centering
  \includegraphics[width=\linewidth]{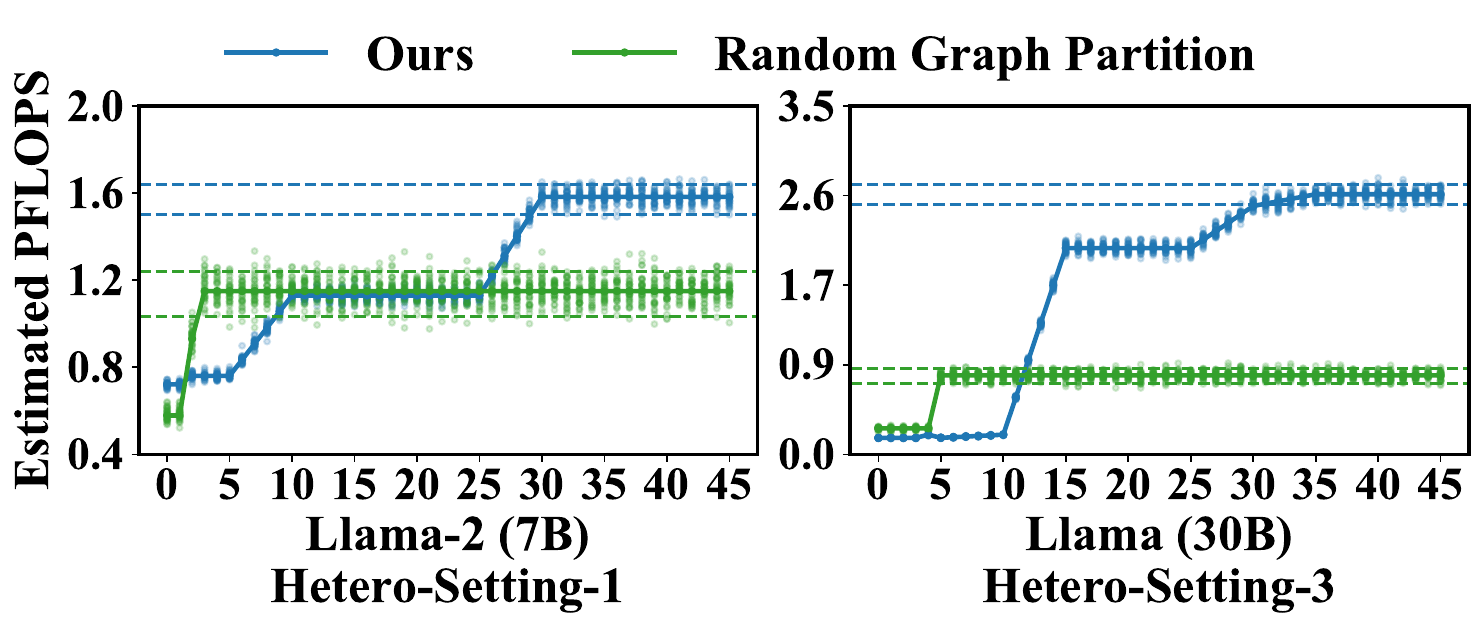}
  \vspace{-1em}
  \caption{Convergence comparison of our algorithm and random graph partition with \textsc{Llama-2 (7B)} (left) and \textsc{(30B)} (right) models, where both run 20 times.}
  \label{fig:search_efficiency}
  \vspace{-1em}
  
\end{figure}

\noindent \textbf{Evaluate the effectiveness of graph partition}. We evaluate our graph partition algorithm by comparing the algorithm convergence gap when running the carefully designed graph partition algorithm and a random graph partition in multiple rounds. As shown in \autoref{fig:search_efficiency}, our algorithm outperforms random graph partition when searching parallel execution plan for \textsc{Llama-2 (7B)} model in heterogeneous setting 1, and \textsc{Llama (30B)} model in heterogeneous setting 3. Despite fluctuations across different rounds, our algorithm converges to $1.4\times$ and $3.3\times$ higher estimated throughput, for \textsc{Llama-2 (7B)} and \textsc{Llama (30B)}. Furthermore, our algorithm continues to improve over iterations, whereas the random graph partitioning method typically converges early without significant improvement. The performance edge of our algorithm becomes more pronounced as the cluster complexity increases and the model size grows. One contributing factor is that random graph partitioning is more likely to result in out-of-memory errors.

\ryan{
\noindent \textbf{Evaluate the effectiveness of batch and layer assignment.} Under heterogeneous setting 1 with the \textsc{Llama (13B)} model, we conduct additional ablation studies comparing our greedy strategy against uniform batch and layer distribution. Disabling the greedy layer distribution leads to an 11.46$\%$ degradation in throughput (from 1.57 PFLOPS to 1.39 PFLOPS), while disabling the greedy batch distribution results in a 9.55$\%$ performance degradation (from 1.57 PFLOPS to 1.42 PFLOPS).}

\begin{figure}
  \centering
  \includegraphics[width=\linewidth]{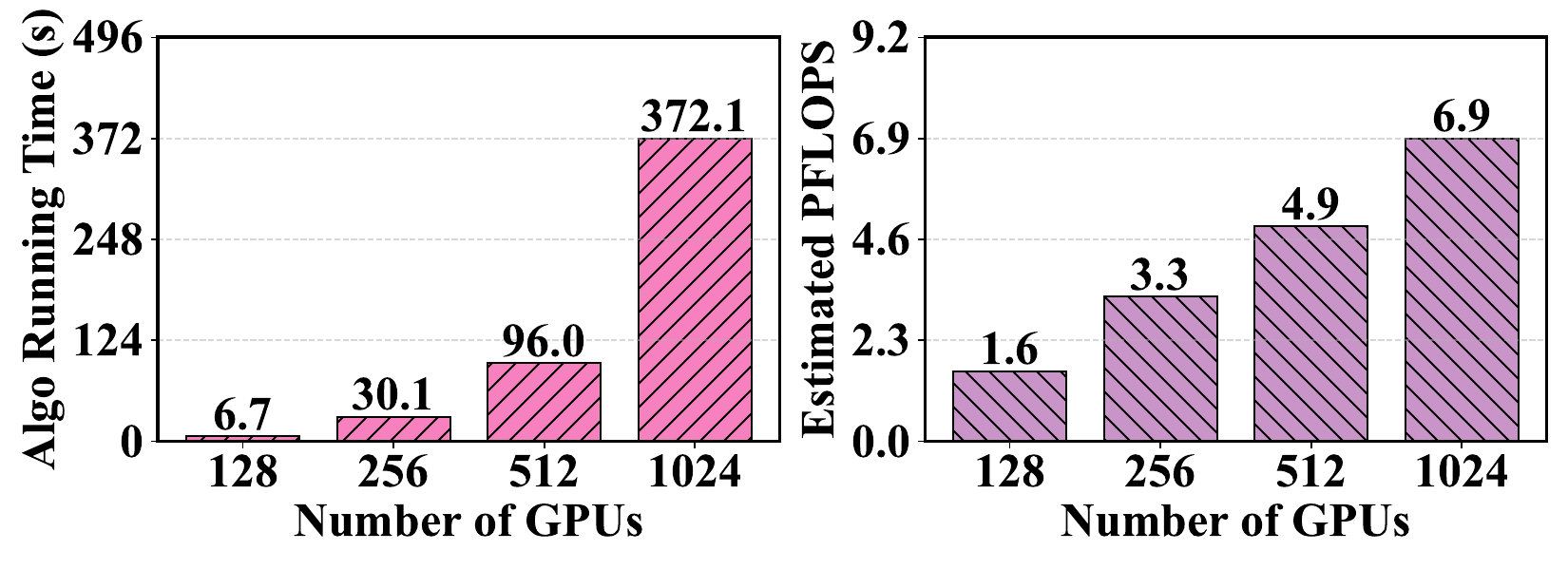}
  \vspace{-1em}
  \caption{Algorithm running time and estimated throughput for GPU cluster sizes (ranging from $4\times$ to $32\times$ GPUs compared to Hetero-Setting 2)}.
  \label{fig:search_speed}
  \vspace{-1em}
\end{figure}

\noindent \textbf{Evaluate the algorithm scalability}. Our graph partition-based scheduling algorithm efficiently handles large graphs, thanks to the introduced heuristics, which prune unnecessary parts of the search space. For example, our coarsening operation (Step (i) in \S\ref{sec:first-phase}) merges graph vertices based on network bandwidth, effectively reducing the complexity of the partitioning process while preserving partition quality. To evaluate the speed of our algorithm, we run $50$ iterations (where our algorithm generally converges) on $12$-core CPUs to search for the optimal parallel execution plan of different numbers of GPUs ($128$ to $1024$). 
As shown in \autoref{fig:search_speed}, our algorithm scales well with an increasing number of GPUs, maintaining a runtime of less than seven minutes even with 1024 GPUs. The scheduling overhead remains manageable, compared to the months-long training time required for large language models. Additionally, the simulated throughput exhibits stable scaling on larger clusters, increasing from 1.6 PFLOPS with 128 GPUs to 6.9 PFLOPS with 1024 GPUs, demonstrating the robustness of our algorithm. 

\begin{figure}[t]
    \centering
    \includegraphics[width=\linewidth]{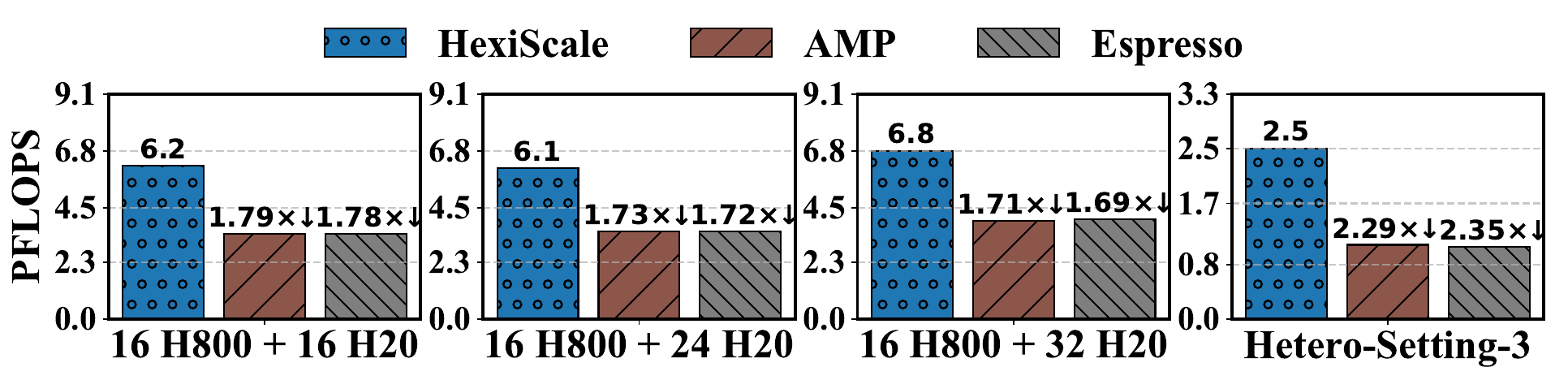}
    \vspace{-1em}
    \caption{Achieved PFLOPS of \sys/AMP/Espresso.}
    \vspace{-1em}
    \label{fig:amp_espresso}

\end{figure}

\begin{figure}[t]
    \centering
    \includegraphics[width=\linewidth]{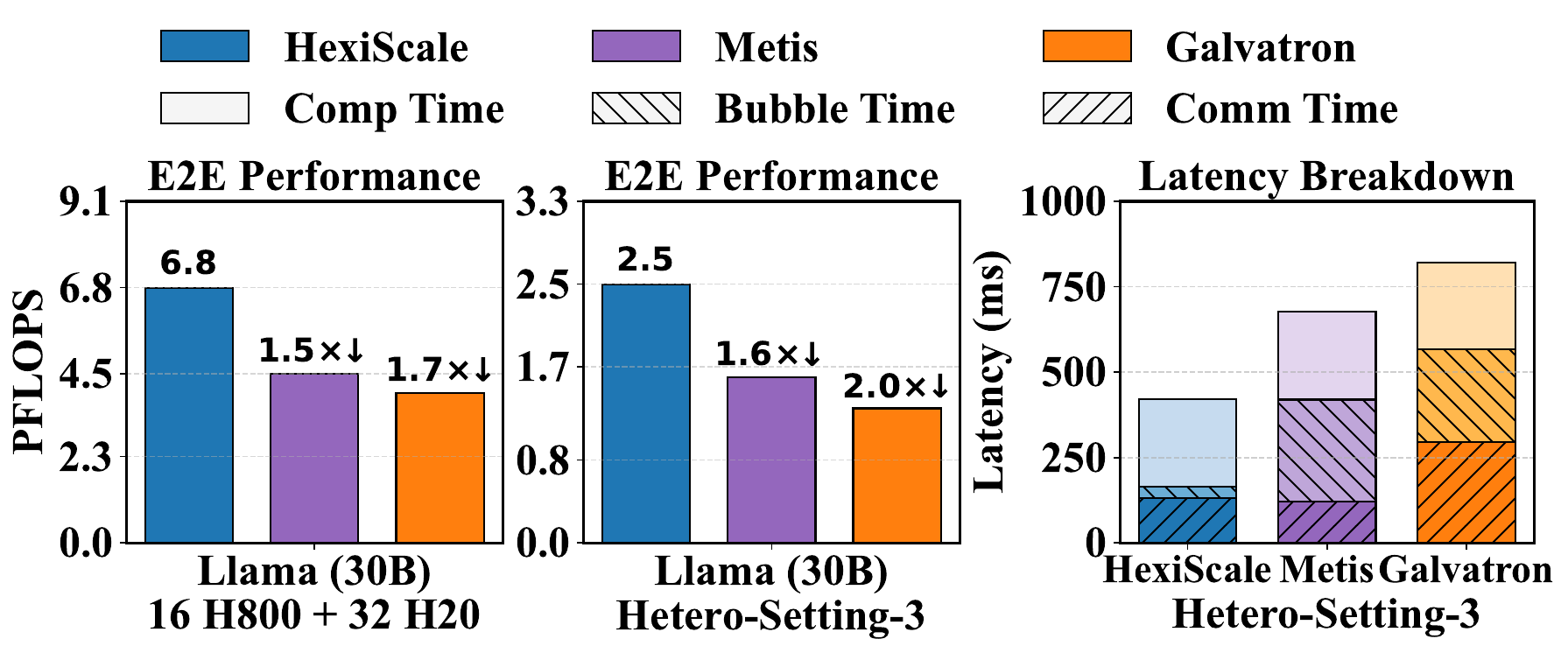}
    
    \vspace{-1em}
    \caption{End-to-end performance comparison (left, middle), and per-batch latency breakdown (right) of three systems. 
    }
    \vspace{-1em}
    \label{fig:e2e_metis}
    
\end{figure}

\subsection{Case Studies}
\label{subsec:simulation}

\noindent \ryan{To further evaluate \sys, we present case studies in this section, comparing \sys with state-of-the-art heterogeneity-aware training systems. We present detailed discussions as follows, and present the detailed experimental configurations for each system in Appendix \ref{app:eval_detail}.}

\noindent \textbf{Compare with AMP/Espresso.} To compare \sys with AMP/Espresso, we conduct experiments for \textsc{Llama (30B)} under heterogeneous settings with both low and high inter-machine network bandwidth. The low-bandwidth scenario is enabled by heterogeneous setting 3; the high-bandwidth scenarios are enabled by three heterogeneous settings with InfiniBand. Compared to AMP/Espresso, results in \autoref{fig:amp_espresso} demonstrate that: (\underline{i}) with low inter-connections, \sys achieves 2.29$\times$/2.35$\times$ higher throughput; (\underline{ii}) with high inter-connections, \sys achieves on average 1.75$\times$/1.73$\times$ higher throughput. In both favorable and unfavorable network conditions, \sys consistently delivers superior performance through flexible asymmetric parallel support and effective scheduling.

\noindent \textbf{Compare with Metis.} To fairly compare \sys with Metis, one of the state-of-the-art heterogeneous training systems~\cite{um2024metis}, we conduct experiments for \textsc{Llama (30B)} on heterogeneous settings with both low (enabled by heterogeneous setting 3) and high (enabled by 16 H800 and 32 H20 with InfiniBand) inter-machine network bandwidth. Since Metis does not open-source its training system and Metis's strategy is deployable via \sys, we run its open-source scheduling algorithm, and deploy the searched parallelism strategy on \sys. Experimental results in \autoref{fig:e2e_metis} (left, middle) demonstrate that Metis achieves a throughput of $4.53$ and $1.56$ PFLOPS on the given two settings, which is $1.5\times$ and $1.6\times$ lower than that of \sys. This provides strong evidence supporting the design of \sys: the limitation of Metis’s algorithmic design (see details in \S\ref{sec:rel}) overlooks opportunities where tensor model parallelism can further enhance system performance. Although tensor model parallelism is more communication-intensive than pipeline parallelism, it incurs no bubble overhead, which can be a potential bottleneck in heterogeneous environments. (\underline{i}) For heterogeneous settings 3, Metis recommends using data parallelism within each machine and pipeline parallelism across the 8 machines (detailed strategy is listed in Appendix \ref{app:eval_detail}). (\underline{ii}) For 16 H800 and 32 H20, Metis suggests using data parallelism within each machine and pipeline parallelism across the 6 machines. In both cases, the long pipeline configuration suggested by Metis incurs significant pipeline bubble overhead. In contrast, \sys demonstrates superior scheduling across different forms of parallelism. Compared to Metis, the key difference lies in the tensor and data parallelism search space: \sys explores $O(D^2)$ combinations, whereas Metis considers only $O(D)$, where $D$ represents the number of GPUs assigned to pipeline stages. For example, tensor model parallelism is additionally incorporated in the strategy suggested by \sys (detailed in \autoref{tab:sys_strategy_refined}) to boost system performance.

\begin{figure}[t]
    \centering
    \includegraphics[width=\linewidth]{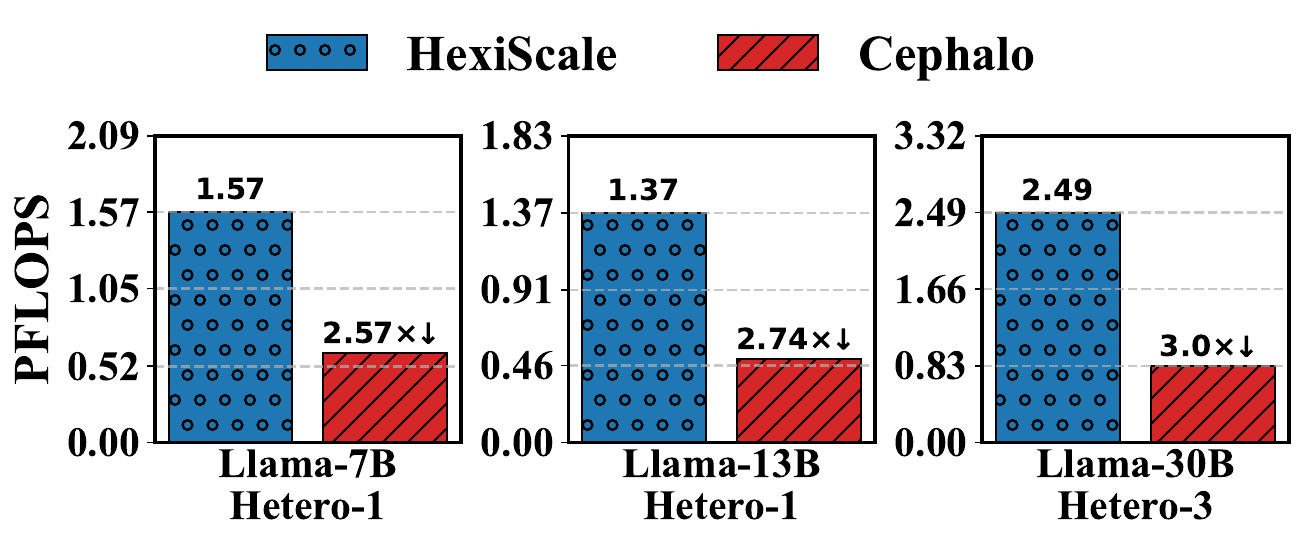}
    \vspace{-1em}
    \caption{Achieved PFLOPS of \sys and Cephalo.}
    \vspace{-1em}
    \label{fig:cephalo}

\end{figure}

\ryan{

\textbf{Compare with Cephalo}. Cephalo's design includes support for uneven assignment of batches and training states (i.e., parameters, gradients, and optimizer states), and optimized gradient accumulation that leverages activation offloading. In contrast, \sys enables fully asymmetric parallel support, i.e., asymmetric pipeline- and data-parallelism. Compared to Cephalo, \sys accommodates stringent network conditions, e.g., when low-speed Ethernet is used for inter-machine communication, and delivers superior performance. Since Cephalo is closed-source, we conduct additional comparisons between \sys and Cephalo through simulations implemented by us over our heterogeneous settings 1 and heterogeneous setting 3 (5.6Gbps inter-machine bandwidth). Experimental results presented in \autoref{fig:cephalo} show that \sys achieves up to 3.0$\times$ higher PFLOPS and, on average, 2.8$\times$ higher PFLOPS than Cephalo. Note that under practical low inter-machine bandwidth, the data-parallel communication overhead in Cephalo significantly degrades performance. In contrast, \sys incurs moderate communication overhead due to its hybrid and asymmetric parallelism design.

}

\noindent \textbf{Latency breakdown comparison.}
For heterogeneous setting 3, we present a detailed breakdown of training iteration latency in \autoref{fig:e2e_metis} (right). Compared to Metis, \sys incurs slightly higher communication overhead, primarily due to its increased demand for tensor model and data parallel communication. However, \sys significantly reduces pipeline bubble time. Metis’s design necessitates a high degree of pipeline parallelism, inevitably increasing bubble time. In contrast, \sys efficiently mitigates this problem by incorporating tensor model parallelism and reducing the number of pipeline stages, thus achieving superior performance.

\noindent \textbf{More experimental results.}
 We compare our scheduling algorithm with an MILP-based scheduling approach in Appendix \ref{app:optimality}. Experimental results show that our algorithm achieves comparable performance to the MILP approach.

\vspace{-0.5em}
\section{Conclusion}
\label{sec:conclude}
\vspace{-0.5em}
\sys is a novel system that enhances LLM training on heterogeneous GPUs by asymmetric parallelism and effective scheduling. Empirical studies show that when running over heterogeneous GPUs, \sys achieves comparable throughput to state-of-the-art homogeneous training systems, and outperforms state-of-the-art heterogeneous training systems with the same heterogeneous GPUs.

\section*{Acknowledgment}

This work is supported by the HKUST startup grant R9895
from CSE; RGC-ECS project 26218024; RGC-NSFC project CRS\_HKUST601/24; Webank-HKUST joint lab project (\textsc{WEB24EG01-E-CSE}. We thank the gift credit from Ucloud that supports this project. We also thank the support from the National Supercomputer Center in Guangzhou Nansha Sub-Center.

\nocite{jianghexgen,jiang2023osdp,jiang2025demystifying,jiang2025hexgen,jiang2025thunderserve,wang2024improving}

\bibliography{example_paper}
\bibliographystyle{mlsys2025}

\clearpage
\appendix

\clearpage
\appendix

\section{Parallelize LLM Training}
\label{sec:para_training}
\noindent\textbf{Parallel training strategies.} To distribute LLM training computation over thousands of GPUs, three main categories of parallel strategies have been proposed.

\textit{Data parallelism}~\cite{li2020pytorch} distributes computation by dividing training batch across devices, where each GPU hosts a model replica for forward and backward propagation, and the gradients are synchronized by \texttt{AllReduce} operations. To further optimize memory usage, gradients, Zero Redundancy Optimizer (ZeRO)~\cite{rajbhandari2020zero} and Fully Sharded Data Parallel (FSDP)~\cite{zhao2023pytorch} further shard the optimizer states and parameters across multiple devices and gather them through additional communication when necessary.

\textit{Pipeline parallelism}~\cite{huang2019gpipe,narayanan2019pipedream} partitions the model's computation across multiple layers into a series of stages, where each GPU handles one stage. During the forward propagation, the GPU serving stage-($j$) sends the activations to the GPU serving stage-($j{+}1$); during the backward pass, communication reverses direction to transfer the gradients. 

\textit{Tensor model parallelism}~\cite{narayanan2021efficient} further partitions each transformer layer over multiple GPUs, where weight matrices are distributed row- or column-wisely. Two \texttt{AllReduce} operations are required to aggregate the layer output activations in forward pass and corresponding gradients in backward pass, respectively. Tensor model parallelism is communication intensive, but can effectively parallelize the computation when the connection between GPUs is fast (e.g., through NVLink).


\noindent\textbf{System optimization for LLM training.} Significant efforts have been made to \textit{optimize parallel LLM training}, and various system optimizations~\cite{park2024accelerating, li2024daha, ai2024neutronorch,jia2019beyond,jia2018exploring,cai2021tensoropt} have been proposed to improve LLM training throughput. For example, Zero bubble parallelism enhances computational efficiency by reducing bubble overhead~\cite{qi2024zero}; A cross-mesh resharding mechanism was introduced to minimize communication overhead in tensor parallelism~\cite{zhuang2023optimizing}; Additionally, system optimizations such as gradient bucketing, gradient accumulation, and computation-communication overlap have been integrated into data parallel implementations~\cite{li2020pytorch}.

To improve training throughput, \textit{optimizations for memory footprint} are also necessary. For example, activation recomputing significantly reduces memory footprint by recomputing the desired activation during backward propagation instead of storing it after forward computation~\cite{chen2024optimizing}. On the other hand, offloading activations to CPU RAM or SSD can effectively reduce GPU memory usage without impacting performance by adaptively overlapping data transfers with computation~\cite{wu2024tba}.

\section{Details of Cost Model}
\label{app:cost_model}
\ryan{
In this section, we present our cost models. We first model the cost for each transformer layer, and then derive the end-to-end cost.

\noindent \textbf{Modeling Cost Layer-wisely}

\begin{itemize}[topsep=5pt, leftmargin=*]

\item \textbf{Tensor model parallel communication cost}. Suppose activation recompute is enabled. A transformer layer runs over a set of GPUs $\mathbf{d}_{i,j}$. The tensor model parallel communication cost for a micro-batch---with a micro-batch size as $B_{mb}$---is:
\begin{small}
\begin{equation}
\begin{aligned}
\textsc{Comm-TP-Fwd}(\mathbf{d}_{i,j}) = \textsc{Comm-TP-Bwd}(\mathbf{d}_{i,j}) \\ =
4 \cdot \max_{d \in \mathbf{d}_{i,j}}
\sum_{d' \in \mathbf{d}_{i,j} \setminus \{d\}}
\left(
\alpha_{d,d'} +
\frac{B_{mb} S H B_{\text{type}}}{|\mathbf{d}_{i,j}| \beta_{d,d'}}
\right)
\end{aligned}
\end{equation}
\end{small}

\item \textbf{Computation cost}. Assume tensor model parallelism runs over the same type of device $d$, for one micro-batch, the computation cost for forward and backward passes is:
\begin{small}
\begin{equation}
\begin{aligned}
\textsc{Comp-TP-Fwd}(\mathbf{d}_{i,j}) =
\frac{24 B_{mb} S H^2 \left(1 + \frac{S}{6H}\right)}
{c_d |\mathbf{d}_{i,j}|} \\
\textsc{Comp-TP-Bwd}(\mathbf{d}_{i,j}) = 2\times \textsc{Comp-TP-Fwd}(\mathbf{d}_{i,j})
\end{aligned}
\end{equation}
\end{small}

\item \textbf{Integrate data parallel communication cost}. We model data parallel strategies (DDP, ZeRO-2, ZeRO-3) as a \textit{per layer overhead} that captures communication and its overlap with computation. For consistency with the tensor-parallel model, we explicitly define the forward and backward computation time of a layer as:
\begin{small}
\begin{equation}
\begin{aligned}
T_f(\mathbf{d}_{i,j}) &= 
\textsc{Comp-TP-Fwd}(\mathbf{d}_{i,j}) 
+ \textsc{Comm-TP-Fwd}(\mathbf{d}_{i,j}) \\
T_b(\mathbf{d}_{i,j}) &= 
\textsc{Comp-TP-Bwd}(\mathbf{d}_{i,j}) 
+ \textsc{Comm-TP-Bwd}(\mathbf{d}_{i,j})
\end{aligned}
\end{equation}
\end{small}

Let $T_{\text{comm}}$ denote the communication cost of ReduceScatter and AllGather, which are data parallel collectives. We model $T_{\text{comm}}$ as follows. Notice that for the AllReduce operation, we model the cost as $2T_{\text{comm}}$ since the AllReduce operation can be implemented via one ReduceScatter and one AllGather operation.
\begin{small}
\begin{equation}
\begin{aligned}
T_{\text{comm}}(k, \sigma) =  \max_{d \in \mathbf{D}^{k}_{\text{dp}}} \sum_{d' \in {\mathbf{D}^{k}_{\text{dp}}} - \{d\}} \left(\alpha_{d, d'} + \frac{12 H^2 B_{\text{type}}} { {\left|\mathbf{D}^{k}_{\text{dp}}\right| \beta}_{d, d'}}\right)
\end{aligned}
\end{equation}
\end{small}

Combining the above formulations, we define the per-layer cost under different data parallel strategies as follows. Since computation-communication overlapping would influence system efficiency (e.g., competing GPU SMs), we introduce constants $\theta_1$, $\theta_2$ for DDP and ZeRO-2’s backward-pass and $\theta_{3f}$, $\theta_{3b}$ for ZeRO-3's forward backward passes that represent the corresponding slowdown. The constants are obtained through profiling.
\begin{small}
\begin{equation}
\begin{aligned}
&\textsc{DP-Overhead-Layer}(k, \sigma) =
\\&\begin{cases}
n_{mb}T_f+\left(n_{mb}-1\right)T_b+\theta_1\cdot\max(T_b,2T_{\text{comm}}), & \text{DDP} \\
n_{mb}T_f+\left(n_{mb}-1\right)T_b+\theta_2\cdot\max(T_b,2T_{\text{comm}}), & \text{ZeRO-2} \\
n_{mb}\theta_{3f}\cdot\max(T_f,T_{\text{comm}}) \\
\quad +\left(n_{mb}-1\right)\theta_{3b}\cdot\max(T_b,T_{comm})& \text{ZeRO-3}\\
\quad +\theta_{3b}\cdot\max(T_b,2T_{comm}) \\
\end{cases}
\end{aligned}
\end{equation}
\end{small}

\noindent \textbf{Explainations for the above formulations}. (\underline{i}) For DDP, which replicates training states, one AllReduce operation is used for gradient communication. The last micro-batch can overlap backward-pass computation with gradient synchronization. (\underline{ii}) For ZeRO-2, which shards gradients and optimizer states, after gradient computation, one ReduceScatter operation is used to synchronize gradients, followed by one AllGather operation to collect updated parameters. The last micro-batch can overlap backward-pass computation with gradients and updated parameters synchronization. (\underline{iii}) For ZeRO-3, which fully shards training states, one AllGather operation is required to collect parameters before the forward-pass and backward-pass computation, and one ReduceScatter operation is used to synchronize gradients. All micro-batches can overlap forward-pass and backward-pass computation with AllGather of parameters, while the last micro-batch can further overlap computation with gradients synchronization.

This formulation ensures that tensor parallel computation and communication are consistently incorporated into data parallel modeling, while avoiding double-counting and explicitly capturing computation--communication overlap.

\item \textbf{Pipeline parallel communication cost}. Pipeline communication occurs between adjacent stages. Define $\textsc{Comm-PP-Hop}(\mathbf{d}_{i,D_{pp}^i}, \mathbf{d}_{i,D_{pp}^i+1}) = 0$. Then:
\begin{small}
\begin{equation}
\begin{aligned}
&\textsc{Comm-PP-Hop}(\mathbf{d}_{i,j}, \mathbf{d}_{i,j+1}) = \\
&2 \cdot \min_{d \in \mathbf{d}_{i,j}, d' \in \mathbf{d}_{i,j+1}}
\Bigg[
\left(
\alpha_{d,d'} +
\frac{B_{mb} S H B_{\text{type}}}{\beta_{d,d'}}
\right) \\
&+ \sum_{d'' \in \mathbf{d}_{i,j+1} \setminus \{d'\}}
\left(
\alpha_{d',d''} +
\frac{B_{mb} S H B_{\text{type}}}
{|\mathbf{d}_{i,j+1}| \beta_{d',d''}}
\right)
\Bigg]
\end{aligned}
\end{equation}
\end{small}

\end{itemize}

\noindent \textbf{Modeling Cost for Each Parallel Strategy}

\begin{itemize}[topsep=5pt, leftmargin=*]

\item \textbf{Pipeline stage cost}. For stage $j$ in pipeline $i$:
\begin{small}
\begin{equation}
\begin{aligned}
\textsc{Stage}(\mathbf{d}_{i,j}) =
\sum_{k=1}^{l_{i,j}}
\Big[
\textsc{DP-Overhead-Layer}(k, \sigma)
\Big]
\end{aligned}
\end{equation}
\end{small}

\item \textbf{Pipeline parallelism cost}. The total execution time for pipeline $i$ is:
\begin{small}
\begin{equation}
\begin{aligned}
\textsc{Pipeline-Time}(i) &=
\sum_{j=1}^{D^i_{pp}}
\Big(
\textsc{Stage}(\mathbf{d}_{i,j}) \\
&+ \textsc{Comm-PP-Hop}(\mathbf{d}_{i,j}, \mathbf{d}_{i,j+1})
\Big) \\
&+ (n_{mb} - 1) \cdot
\max_{j=2,\dots,D^i_{pp}}
\Big(
\textsc{Stage}(\mathbf{d}_{i,j}) \\
&+ \textsc{Comm-PP-Hop}(\mathbf{d}_{i,j}, \mathbf{d}_{i,j+1})
\Big)
\end{aligned}
\end{equation}
\end{small}

\end{itemize}

\noindent \textbf{Modeling end-to-end time} The iteration time is determined by the slowest pipeline:

\begin{small}
\begin{equation}
\begin{aligned}
\textsc{Total-Time}(\sigma) =
\max_{i=1,\dots,D_{dp}}
\left\{ \textsc{Pipeline-Time}(i) \right\}
\end{aligned}
\end{equation}
\end{small}

\noindent \textbf{Modeling Memory Cost:} Suppose full activation recompute and distributed data parallelism (DDP) are applied. The memory cost of parameters and activations for one layer can be estimated as follows:
\begin{small}
\begin{equation}
\begin{aligned}
    \textsc{Mem-Layer} \left( \sigma \right)= \frac{48H^2B_{type}}{|\mathbf{d}_{i,j}|} + B_{mb}SHB_{type}
\end{aligned}
\end{equation}
\end{small}

\ryan{Under the 1F1B schedule, the peak memory is given by:
\begin{equation}
\textsc{Mem-Peak}\left( \sigma \right) = L \cdot \textsc{Mem-Layer}\left( \sigma \right).
\end{equation}}
}

\section{Detailed Validations on Cost Model}
\label{app:cost_model_val}

\noindent \textbf{Evaluate the simulation accuracy}. We evaluate the accuracy of our simulation in \autoref{tab:simulation-comp}. The simulation results (deviating less than $2.9\%$) closely match the actual outcomes across various GPU settings. Our algorithm can thereby accurately search for an effective parallel execution plan. To further validate our cost model, we present additional experiments in Appendix~\ref{app:cost_model_val}. The results demonstrate that our cost model closely predicts both memory usage and per-batch communication overhead.

\begin{table}[t]
\centering
\caption{Comparison of the real and simulated achieved PFLOPS across different experimental settings.}
\label{tab:simulation-comp}
\resizebox{\linewidth}{!}{
\begin{tabular}{ l | l | c c | c }
\toprule
 \multirow{2}{*}{\textbf{Model}} & \multirow{2}{*}{\textbf{Setting}} & \textbf{Real} & \textbf{Simulated} & \textbf{Deviation} \\ 
   &  & \textbf{(PFLOPS)} & \textbf{(PFLOPS)} & \textbf{(\%)} \\ 
 \midrule
\multirow{4}{*}{\textsc{Llama-2 (7B)}} 
 & Homo-Ethernet     & $2.05$ & $2.11$ & $2.84\%$ \\ 
 & Homo-RDMA         & $2.68$ & $2.70$ & $0.74\%$ \\ 
 & Hetero-Setting-1  & $1.57$ & $1.61$ & $2.48\%$ \\ 
 & Hetero-Setting-2  & $1.65$ & $1.66$ & $0.60\%$ \\ \midrule

\multirow{4}{*}{\textsc{Llama-2 (13B)}} 
 & Homo-Ethernet     & $1.97$ & $2.02$ & $2.48\%$ \\ 
 & Homo-RDMA         & $2.55$ & $2.61$ & $2.30\%$ \\ 
 & Hetero-Setting-1  & $1.37$ & $1.40$ & $2.14\%$ \\ 
 & Hetero-Setting-2  & $1.54$ & $1.56$ & $1.28\%$ \\ \midrule

\multirow{3}{*}{\textsc{Llama (30B)}} 
 & Homo-Ethernet     & $2.78$ & $2.86$ & $2.80\%$ \\ 
 & Homo-RDMA         & $3.89$ & $3.99$ & $2.51\%$ \\ 
 & Hetero-Setting-3  & $2.49$ & $2.54$ & $1.97\%$ \\ \bottomrule
\end{tabular}
}
\end{table}

Comparisons on predicted and real memory costs (see \ref{tab:peak_memory}) and communication costs (see \ref{tab:comm_latency}) across different models validate our cost model. The predicted values closely align with the observed values during real execution.

\begin{table}[t]
\centering
\caption{Estimated vs. Observed Peak Memory Usage.}
\label{tab:peak_memory}
\begin{tabular}{l|c|c}
\toprule
\multirow{2}{*}{\textbf{Model}} & \textbf{Estimated Peak} & \textbf{Observed Peak} \\
 & \textbf{Memory (GB)} & \textbf{Memory (GB)} \\
\midrule
\textsc{Llama-2 (7B)}  & 48.2  & 52.3  \\
\textsc{Llama-2 (13B)} & 95.8  & 98.4  \\
\textsc{Llama (30B)} & 239.4 & 246.2 \\
\bottomrule
\end{tabular}
\end{table}

\begin{table}[t]
\centering
\caption{Estimated vs. Observed Communication Latency in Heterogeneous Settings. (following \autoref{fig:lat_bk})}
\label{tab:comm_latency}
\begin{tabular}{l|c|c}
\toprule
\multirow{2}{*}{\textbf{Model}} & \textbf{Estimated} & \textbf{Observed} \\
 & \textbf{Latency (ms)} & \textbf{Latency (ms)} \\
\midrule
\textsc{Llama-2 (7B)} & 32.31  & 32.66  \\
\textsc{Llama-2 (13B)} & 103.24 & 104.28 \\
\textsc{Llama (30B)} & 129.27 & 130.51 \\
\bottomrule
\end{tabular}
\end{table}



\section{Details of Experiment Setups}
\label{app:exp_setup}

We simulate the performance of \sys, Metis, and Galvatron on a
large-scale cluster consisting of 240 heterogeneous GPUs, the results are summarized in \autoref{tab:large_scale_gpu_config}:

\begin{table}[ht]
    \centering
    \caption{GPU compositions in the simulated large-scale cluster.}
    \label{tab:large_scale_gpu_config}
    \begin{tabular}{lcc}
        \toprule
        \textbf{GPU Type} & \textbf{Num Instances} & \textbf{Bandwidth (GB/s)} \\
        \midrule
        \texttt{8$\times$3090}  & 5  & 24  \\
        \texttt{4$\times$3090}  & 6  & 24  \\
        \texttt{8$\times$4090}  & 20 & 32  \\
        \texttt{8$\times$A800}  & 2  & 200 \\
        \bottomrule
    \end{tabular}
\end{table}

\section{Algorithm Optimality}
\label{app:optimality}

Due to the NP-hard nature of this complex problem, determining an upper bound is particularly challenging, as it requires either a provably optimal solution or a heuristic guarantee under specific constraints. To evaluate the optimality of our scheduling algorithm, we compare it against the MILP (Mixed-Integer Linear Programming) approach, a widely adopted method in resource allocation problems. ILP/MILP-based solvers have been integrated into prominent systems such as Alpa~\cite{zheng2022alpa}, FlexSP~\cite{wang2024data}, and Helix~\cite{mei2024helix}, demonstrating their reliability and effectiveness in resource allocation and parallel strategy determination for LLM training and inference.
    
We run both a MILP algorithm and our algorithm, and compare the performance gap. The MILP algorithm guarantees optimal results, except its running time grows exponentially.
    
In \autoref{tab:milp}, we simulate the system performance for heterogeneous settings that appear in \S \ref{subsec:e2e}. As shown in the table, the scheduling algorithm of \sys achieves performance comparable to MILP. Additionally, the graph partitioning-based scheduling algorithm used by \sys excels in scalability for complex and large environments, significantly outperforming MILP in terms of search time (\sys completes the search within minutes, whereas MILP requires hours). In summary, our algorithm delivers performance close to the optimal solution while demonstrating exceptional efficiency.

\begin{table}[h]
    \centering
    \caption{Throughput comparison of MILP and \sys across different models and heterogeneous settings.}
    \label{tab:milp}
    \resizebox{\linewidth}{!}{ 
    \begin{tabular}{l l l c}
        \toprule
        \textbf{Model} & \textbf{Setting} & \textbf{Method} & \textbf{Achieved PFLOPS} \\
        \midrule
        \multirow{2}{*}{Llama-2 (7B)}  & \multirow{2}{*}{Hetero-Setting 1} & MILP     & 1.62 \\
                                       &                                   & \sys     & 1.62 \\
        \multirow{2}{*}{Llama-2 (13B)} & \multirow{2}{*}{Hetero-Setting 1} & MILP     & 1.46 \\
                                       &                                   & \sys     & 1.44 \\
        \multirow{2}{*}{Llama-2 (7B)}  & \multirow{2}{*}{Hetero-Setting 2} & MILP     & 1.70 \\
                                       &                                   & \sys     & 1.66 \\
        \multirow{2}{*}{Llama-2 (13B)} & \multirow{2}{*}{Hetero-Setting 2} & MILP     & 1.63 \\
                                       &                                   & \sys     & 1.56 \\
        \multirow{2}{*}{Llama (30B)}   & \multirow{2}{*}{Hetero-Setting 3} & MILP     & 2.69 \\
                                       &                                   & \sys     & 2.54 \\
        \bottomrule
    \end{tabular}
    }
\end{table}

\section{Simulation Experiments}

\begin{figure}[t]
    \centering
    \includegraphics[width=\linewidth]{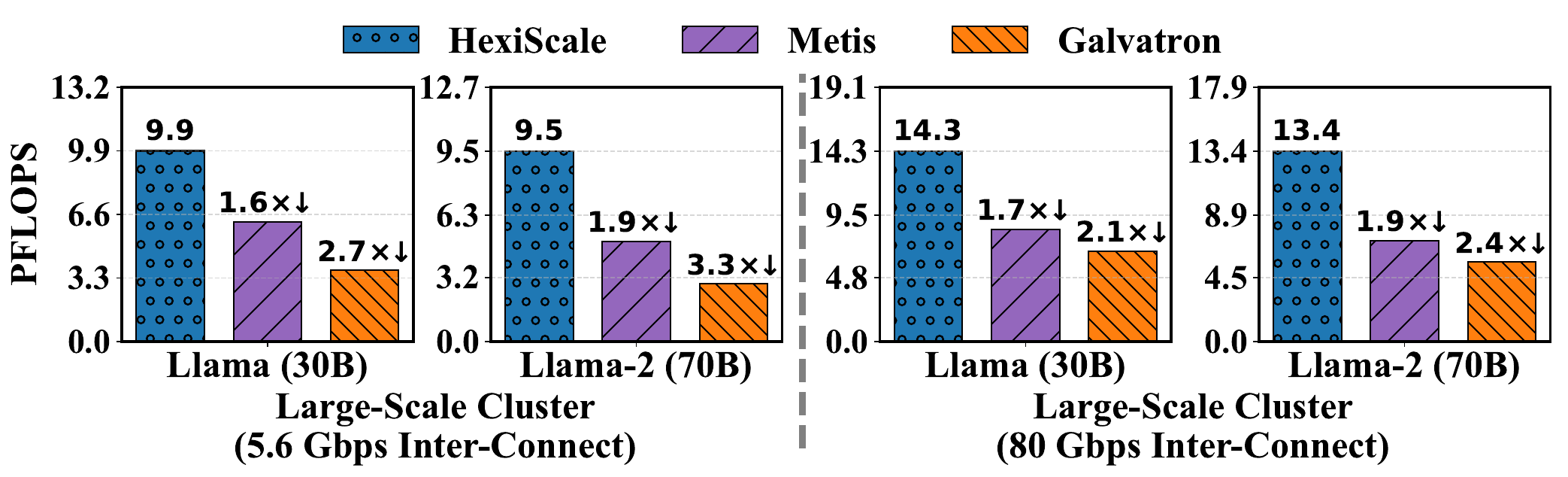}
    
    \vspace{-1em}
    \caption{Large scale simulation of \sys, Metis, and Galvatron on \textsc{Llama (30B)} and \textsc{Llama-2 (70B)} models.}
    
    \label{fig:large_scale_sim}
\end{figure}

\noindent \textbf{Large-scale simulation.} To further evaluate \sys on larger models and cluster configurations, we simulate performance of \sys, Metis, and Galvatron on a 240 heterogeneous GPU large-scale cluster with \textsc{Llama (30B)} and \textsc{Llama-2 (70B)} under both low (5.6 Gbps) and high (80 Gbps) inter-machine network bandwidth. Detailed cluster information is provided in Appendix \ref{app:exp_setup}. As illustrated in~\autoref{fig:large_scale_sim}, under both low and high inter-machine network connections, \sys exhibits superior performance on the large-scale cluster with different model sizes, achieving performance improvements of 1.6$\times$ to 1.9$\times$ relative to Metis, and 2.1$\times$ to 3.3$\times$ relative to Galvatron. Additionally, Metis's and Galvatron's search algorithms fail to efficiently identify high-performance parallelism strategies in large-scale clusters, requiring hours to complete. In contrast, \sys demonstrates superior algorithmic scalability, as discussed in \S\ref{subsec:algo}.

\begin{figure}[t]
    \centering
    \includegraphics[width=\linewidth]{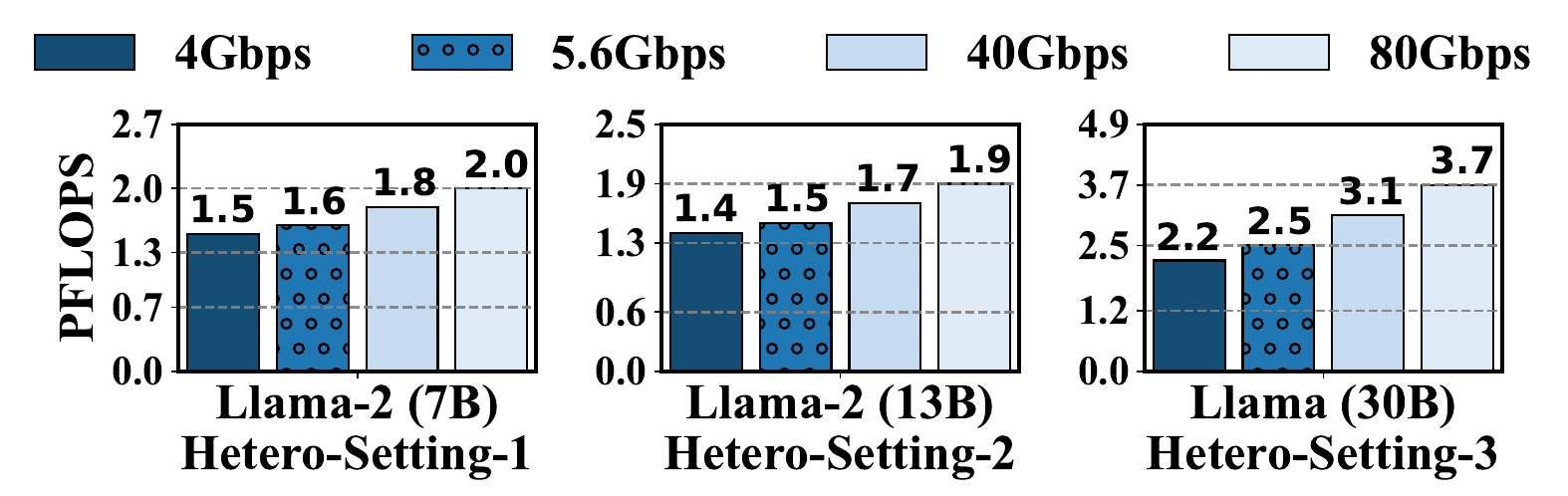}
    \vspace{-1em}
    \caption{Simulated throughput achieved by \sys under varied inter-machine network connections. 
    }
    
    \label{fig:net_var}
    
\end{figure}

\noindent \textbf{Performance under varied network.} Network connectivity plays a critical role in heterogeneous training. To evaluate how \sys performs under varying network conditions, we simulate system performance by adjusting inter-machine bandwidths---from 4 Gbps to 80 Gbps---for \textsc{LLaMA-2 (7B)}, \textsc{LLaMA-2 (13B)}, and \textsc{LLaMA (30B)} models in heterogeneous settings described in \S\ref{subsec:e2e}. The simulation results are summarized in \autoref{fig:net_var}. When inter-machine bandwidth degrades from 5.6 Gbps to 4 Gbps, \sys maintains robust performance with only moderate degradation. Conversely, as bandwidth increases to 80 Gbps, \sys achieves a notable boost in system performance. At 80 Gbps, \sys delivers throughput comparable to the homogeneous baseline with RDMA—up to $0.95\times$, and $0.80\times$ on average—despite operating in a heterogeneous environment. Importantly, upgrading the network fabric (e.g., enabling RoCE at 40 Gbps or RDMA at 80 Gbps) is significantly more cost-effective than replacing GPUs, suggesting that heterogeneous training—enabled by \sys---offers a promising and economical alternative to conventional homogeneous setups.

\ryan{
\section{Evaluation Configuration Details}
\label{app:eval_detail}

\begin{table}[t]
\caption{Parallelism strategy discovered by \sys when training \textsc{Llama (30B)} in heterogeneous setting 3. There are four pipelines (DP=4) with various pipeline layouts.}
\centering
\resizebox{\linewidth}{!}{
\begin{tabular}{c | c | c | c | c}
\toprule
\textbf{Pipeline} & \textbf{Stage} & \textbf{GPU} & \textbf{Layer} & \textbf{TP} \\
\textbf{Index}    & \textbf{Index} & \textbf{Allocation} & \textbf{Count} & \textbf{Degree} \\
\midrule
0 & \{0\}             & 8$\times$A800   & 60 & 8 \\
\midrule
\multirow{2}{*}{1} & \{0,1,2\}       & 4$\times$4090   & 18 & 4 \\
\cmidrule{2-5}
                   & \{3\}          & 2$\times$3090   & 6  & 2 \\
\midrule
\multirow{2}{*}{2} & \{0,1,2\}       & 4$\times$4090   & 18 & 4 \\
\cmidrule{2-5}
                   & \{3\}          & 2$\times$3090   & 6  & 2 \\
\midrule
\multirow{2}{*}{3} & \{0,1\}       & 4$\times$4090   & 18 & 4 \\
\cmidrule{2-5}
                   & \{2,3,4,5,6,7\}      & 2$\times$3090   & 4  & 2 \\
\bottomrule
\end{tabular}%
}
\vspace{-1.5em}
\label{tab:sys_strategy_refined}
\end{table}

\subsection{Case study}

In this subsection, we present the case study of scheduling in the heterogeneous setting 3 to compare the parallel strategy across \sys, Metis, and Galvatron

\noindent \textbf{Strategy of \sys.} \autoref{tab:sys_strategy_refined} illustrates the parallel strategy discovered by our algorithm. Within each pipeline, distinct hardware characteristics are effectively considered by applying locally high-performance parallel strategies and strategically assigning transformer layers to balance computation across pipeline stages. The number of pipelines is fine-tuned to four, leveraging the benefits of data parallelism while avoiding out-of-memory issues and too much communication overhead. Data parallel communication overhead remains manageable, as pipeline execution time is much longer. These results confirm that \sys effectively accounts for hardware heterogeneity, generating high-performance parallel strategies.

The parallel strategies for Metis and Galvatron are as follows:

\noindent \textbf{Strategy of Metis.} Metis prefers using intra-machine data parallelism, thereby constructing one pipeline with 8 stages; the results are summarized in \autoref{tab:metis_strategy}.


\begin{table}[ht]
\caption{Parallel Strategy of Metis under Heterogeneous Setting 3.}
\label{tab:metis_strategy}
\centering
\resizebox{0.8\linewidth}{!}{
\begin{tabular}{c|c|c|c}
\toprule
\textbf{Stage} & \textbf{GPU} & \textbf{Layer} & \textbf{(DP, TP)} \\
\textbf{Index} & \textbf{Allocation} & \textbf{Count} & \textbf{Degree} \\
\midrule
{0} & 8xA800 & 20 & (8, 1) \\
\midrule
{1,2,3,4} & 8x4090 & 8 & (8, 1)  \\
\midrule
{5} & 8x3090 & 4 & (8, 1) \\
\midrule
{6} & 4x3090 & 2 & (4, 1) \\
\midrule
{7} & 4x3090 & 2 & (4, 1) \\
\bottomrule
\end{tabular}
}
\end{table}

\noindent \textbf{Strategy of Galvatron.} The strategy for Galvatron is fine-tuned as $(D_{dp}, D_{tp}, D_{pp})=(4, 2, 7)$. With a higher data parallel degree, the out-of-memory issue persists. With a lower data parallel degree, the system performance is compromised by either significant pipeline bubbles or communication overhead. Tensor model parallel degree is fine-tuned to 2; otherwise, for higher tensor model parallel degrees, 3090 incurs significant communication overhead and compromises the overall system performance.

\subsection{Other Evaluation Details}

\textbf{Pipeline schedules.} For all systems, we adopt 1F1B when enabling pipeline parallelism.

\textbf{Data-parallel sharding strategies.} Megatron/Metis uses DDP, Galvatron/AMP/Espresso employs ZeRO-3. \sys adopts ZeRO-3 for heterogeneous clusters comprising Hopper GPUs, and adopts ZeRO-2 for other settings.

\textbf{FSDP configurations.} FSDP uses ZeRO-3 when RDMA is available and ZeRO-2 when only Ethernet is available. We do not enable activation offloading for FSDP, since empirically it degrades performance on A800 GPUs.

\textbf{Batch size selection criteria.} We tune the global-batch and micro-batch size that leads to the highest throughput for each system.

\textbf{Parallel Strategy and batch size details.} For each system, we summarize the parallel strategy and batch size used in our evaluations in \autoref{tab:hexiscale_metis_30b}, \autoref{tab:main_systems}, \autoref{tab:hexiscale_7b_13b}, and \autoref{tab:hexiscale_hopper}.

\begin{table}[ht]
\caption{Batch Configuration for \sys and Metis on 30B (Heterogeneous Setting 3).}
\label{tab:hexiscale_metis_30b}
\centering
\ryan{
\resizebox{\linewidth}{!}{
\begin{tabular}{c|c|c}
\toprule
\textbf{System} & \textbf{Batch per Pipeline} & \textbf{Micro Batch} \\
\midrule
\sys (4 pipelines) & 80,64,64,48 & 2 \\
Metis (1 pipeline) & 256 & 8 \\
\bottomrule
\end{tabular}
}
}
\end{table}

\begin{table*}[ht]
\caption{Parallelism Strategy and Batch Sizes for Megatron, Galvatron, FSDP, and AMP/Espresso.}
\label{tab:main_systems}
\centering
\ryan{
\resizebox{\linewidth}{!}{
\begin{tabular}{c|c|c|c|c|c}
\toprule
\textbf{System} & \textbf{Model} & \textbf{Cluster} & \textbf{(DP, TP, PP)} & \textbf{Global Batch} & \textbf{Micro Batch} \\
\midrule
Megatron & Llama-7B  & 2x8xA800 & (8,1,2) & 384 & 4 \\
Megatron & Llama-13B & 2x8xA800 & (8,1,2) & 192 & 2 \\
Megatron & Llama-30B & 4x8xA800 & (4,4,2) & 192 & 2 \\
Megatron & Llama-7B  & Hetero Setting 1 & (5,1,8) & 120 & 1 \\
Megatron & Llama-13B & Hetero Setting 1 & (2,1,20) & 128 & 1 \\
Megatron & Llama-7B  & Hetero Setting 2 & (2,1,16) & 128 & 1 \\
Megatron & Llama-13B & Hetero Setting 2 & (2,1,16) & 128 & 1 \\
\midrule
Galvatron & Llama-7B  & 2x8xA800 & (8,1,2) & 384 & 4 \\
Galvatron & Llama-13B & 2x8xA800 & (8,1,2) & 192 & 2 \\
Galvatron & Llama-30B & 4x8xA800 & (4,4,2) & 192 & 2 \\
Galvatron & Llama-7B  & Hetero Setting 1 & (8,1,5) & 128 & 2 \\
Galvatron & Llama-13B & Hetero Setting 1 & (4,1,10) & 128 & 2 \\
Galvatron & Llama-7B  & Hetero Setting 2 & (8,1,4) & 128 & 2 \\
Galvatron & Llama-13B & Hetero Setting 2 & (8,1,4) & 128 & 2 \\
Galvatron & Llama-30B & Hetero Setting 3 & (7,1,8) & 96 & 2 \\
\midrule
FSDP & Llama-7B  & 2x8xA800 & (16,1,1) & 384 & 4 \\
FSDP & Llama-13B & 2x8xA800 & (16,1,1) & 192 & 4 \\
FSDP & Llama-30B & 4x8xA800 & (32,1,1) & 192 & 2 \\
\midrule
AMP/Espresso & Llama-30B & 16xH800+16xH20 & (2,4,4) & 384 & 4 \\
AMP/Espresso & Llama-30B & 16xH800+24xH20 & (2,4,5) & 384 & 4 \\
AMP/Espresso & Llama-30B & 16xH800+32xH20 & (2,4,6) & 384 & 4 \\
AMP/Espresso & Llama-30B & Hetero Setting 3 & (7,1,8) & 128 & 2 \\
\bottomrule
\end{tabular}
}
}
\end{table*}

\begin{table*}[ht]
\caption{\sys Parallel Strategy for 7B/13B Models.}
\label{tab:hexiscale_7b_13b}
\centering
\ryan{
\resizebox{\linewidth}{!}{
\begin{tabular}{c|c|c|c|c|c|c}
\toprule

\textbf{Model} & \textbf{Cluster} & \textbf{\#Pipelines} & \textbf{TP per Stage} & \textbf{GPU Allocation} & \textbf{Global Batch} & \textbf{Micro Batch} \\
\midrule
Llama-7B  & Hetero Setting 1 & 8 & 1,1,2,1 & 1x3090+1x3080+3x4090 & 576 & 4 \\
Llama-13B & Hetero Setting 1 & 4 & 1,1,1,1,2,2,2 & 2x3090+2x3080+6x4090 & 384 & 2 \\
Llama-7B  & Hetero Setting 2 & 8 & 1,1,1,1 & 1x3090+1x3080+1x4090+1xA800 & 576 & 4 \\
Llama-13B & Hetero Setting 2 & 8 & 1,1,1,1 & 1x3090+1x3080+1x4090+1xA800 & 384 & 2 \\
\bottomrule
\end{tabular}
}
}
\end{table*}

\begin{table*}[ht]
\caption{\sys Strategy for 30B on Hopper Heterogeneous Clusters.}
\label{tab:hexiscale_hopper}
\centering
\ryan{
\resizebox{0.9\linewidth}{!}{
\begin{tabular}{c|c|c|c}
\toprule
\textbf{Cluster} & \textbf{Parallel Strategy} & \textbf{Batch per Pipeline} & \textbf{Micro Batch} \\
\midrule
16xH800+16xH20 & P0-1: 8xH800; P2-3: 8xH20 & 160,160,32,32 & 4 \\
\midrule
16xH800+24xH20 & P0-1: 8xH800; P2-4: 8xH20 & 160,160,32,32,32 & 4 \\
\midrule
16xH800+32xH20 & P0-1: 8xH800; P2-5: 8xH20 & 160,160,32,32,32,32 & 4 \\
\bottomrule

\end{tabular}
}
}
\end{table*}

}


\end{document}